\documentclass[11pt,a4paper]{article}

\usepackage[utf8]{inputenc}
\usepackage[T1]{fontenc}
\usepackage{lmodern}
\usepackage[margin=2.5cm]{geometry}
\usepackage{graphicx}
\usepackage{booktabs}
\usepackage{amsmath,amssymb}
\usepackage{enumitem}
\usepackage{hyperref}
\usepackage[super,sort&compress]{natbib}
\usepackage{xcolor}
\usepackage{caption}
\usepackage{subcaption}
\usepackage{tikz}
\usepackage{fontawesome5}
\usepackage{array}
\usepackage{longtable}
\usepackage{xltabular}
\usetikzlibrary{positioning,arrows.meta,shapes.geometric,shapes.symbols,fit,backgrounds,calc}

\graphicspath{{figures/}}
\definecolor{cfrule}{HTML}{8A8A8A}
\definecolor{cfSite}{HTML}{DCE5F1}
\definecolor{cfMain}{HTML}{EAEBED}
\definecolor{cfFinal}{HTML}{DDE8D3}
\definecolor{cfNote}{HTML}{F5F0E4}

\DeclareCaptionLabelSeparator{pipe}{\ $|$\ }
\title{%
  ARMD-Predict: multi-site machine learning to\\
  predict antibiotic resistance before culture results%
}

\author{%
\parbox{\textwidth}{\centering
  Fateme Nateghi Haredasht\textsuperscript{1,*,$\dagger$}, Nicholas P. Marshall\textsuperscript{2,*}, Wenyuan Chen\textsuperscript{1}, Lauren N. Cooper\textsuperscript{3}, Ziming Wei\textsuperscript{4}, Harrison Konsker\textsuperscript{1}, Jacob S. Chen\textsuperscript{1}, Michael Klompas\textsuperscript{4,5}, Cara McKenna\textsuperscript{4,5}, Richard J. Medford\textsuperscript{3,6}, Sanjat Kanjilal\textsuperscript{4}, Austin J. Schoeffler\textsuperscript{7}, Stephen P. Ma\textsuperscript{8}, Niaz Banaei\textsuperscript{9,10}, Steven M. Asch\textsuperscript{11}, Amy Chang\textsuperscript{9}, Stanley C. Deresinski\textsuperscript{9}, Mary K. Goldstein\textsuperscript{12}, Robert Tibshirani\textsuperscript{13}, Jonathan H. Chen\textsuperscript{1,8,14}
  \\[10pt]
  {\footnotesize
  \textsuperscript{1}Division of Computational Medicine, Stanford University, Stanford, CA, USA; \textsuperscript{2}Division of Pediatric Infectious Diseases, Department of Pediatrics, Stanford University School of Medicine, Palo Alto, CA, USA; \textsuperscript{3}Clinical Informatics Center, University of Texas Southwestern Medical Center, Dallas, TX, USA; \textsuperscript{4}Department of Population Medicine, Harvard Medical School and Harvard Pilgrim Health Care Institute, Boston, MA, USA; \textsuperscript{5}Division of Infectious Diseases, Department of Medicine, Brigham and Women's Hospital, Boston, MA, USA; \textsuperscript{6}Brody School of Medicine, Department of Internal Medicine, East Carolina University, Greenville, NC, USA; \textsuperscript{7}Department of Emergency Medicine, Department of Medicine, Stanford University School of Medicine, Stanford, CA, USA; \textsuperscript{8}Division of Hospital Medicine, Department of Medicine, Stanford University School of Medicine, Stanford, CA, USA; \textsuperscript{9}Division of Infectious Diseases and Geographic Medicine, Department of Medicine, Stanford University School of Medicine, Stanford, CA, USA; \textsuperscript{10}Department of Pathology, Stanford University School of Medicine, Stanford, CA, USA; \textsuperscript{11}Division of Primary Care and Population Health, Department of Medicine, Stanford University School of Medicine, Stanford, CA, USA; \textsuperscript{12}Department of Health Policy, Stanford University School of Medicine, Stanford, CA, USA; \textsuperscript{13}Department of Biomedical Data Science and Department of Statistics, Stanford University, Stanford, CA, USA; \textsuperscript{14}Clinical Excellence Research Center, Stanford University School of Medicine, Stanford, CA, USA.
  \\[6pt]
  \textsuperscript{*}These authors contributed equally.\quad
  \textsuperscript{$\dagger$}Corresponding author: \texttt{fnateghi@stanford.edu}}
}}

\date{}

\begin{document}

\maketitle

\begin{abstract}
Clinicians select empiric antibiotics before culture susceptibility results are available. Electronic health record (EHR) models could inform that choice, but most are developed at a single hospital and transport poorly to others. We developed ARMD-Predict using 7.5 million culture--antibiotic observations from four academic medical centers. One set of models estimates susceptibility to individual antibiotics; a second, hierarchical set predicts the infecting organism and then organism-specific resistance phenotypes. Across 23 pooled susceptibility models evaluated in same-period, patient-grouped test partitions, median area under the receiver operating characteristic curve (AUROC) was 0.77 for inpatient urine and 0.70 for inpatient blood cultures; chronological retraining reduced AUROC by approximately 0.04. At an exploratory threshold with 95\% positive predictive value, derived and evaluated on the same test observations, the model identified 55\% of ceftriaxone-susceptible inpatient urine observations as suitable for narrower therapy and flagged 60\% of ceftriaxone-resistant observations for broader coverage. Organism classifiers achieved top-3 accuracy of 51.3--83.3\%; phenotype AUROCs were conditional on the observed organism. Research applications are available at \url{https://armd-predict.stanford.edu}. Independent calibration, threshold and end-to-end prospective validation are required before clinical use.
\end{abstract}

\section{Introduction}\label{sec:introduction}

Antimicrobial resistance was associated with 4.71 million deaths in 2021, including 1.14 million directly attributable deaths, and attributable mortality is projected to reach 1.91 million by 2050 (ref.~\citenum{gbd2024amr}). Clinicians must often select empiric therapy 24 to 72 hours before culture and susceptibility results are available. During this interval, undertreatment exposes patients with resistant infections to inactive therapy and the risk of consequent clinical deterioration, whereas unnecessary broad-spectrum treatment increases selection pressure and may increase risk for adverse effects.

In a meta-analysis of severe infection, appropriate empiric therapy was associated with lower 30-day mortality (relative risk, 0.71; 95\% confidence interval (CI), 0.62--0.82) and in-hospital mortality (relative risk, 0.67; 95\% CI, 0.56--0.80)\cite{marquet2015inappropriate}. Population antibiograms do not account for variation among patients within the same hospital. For example, \emph{Escherichia coli} susceptibility to ciprofloxacin ranged from 64.5\% to 95.1\% across units in one study, illustrating how hospital-level estimates can overestimate or underestimate coverage for an individual patient\cite{kuster2008stratification,xie2025antibiogram}. Clinical scores for methicillin-resistant \emph{Staphylococcus aureus} (MRSA) pneumonia and decision trees for extended-spectrum \ensuremath{\beta}-lactamase (ESBL) bacteremia use patient history, but provide categorical rather than antibiotic-specific estimates\cite{shorr2013risk,goodman2016clinical}.

EHR models can integrate longitudinal history and produce patient-specific organism and susceptibility probabilities at the time of prescribing. Prior studies predicted susceptibility in urinary tract infection and supported outpatient prescribing decisions\cite{yelin2019personal,kanjilal2020decision}. Personalized antibiograms preserved or improved coverage relative to clinician-selected therapy across two academic systems\cite{corbin2020personalized,corbin2022personalized,haredasht2024enhancing,amrollahi2024susceptibility}. In another study, EHR models achieved AUROCs of 0.73--0.79 without organism identity and 0.80--0.88 when organism identity was included\cite{lewinepstein2021predicting}. Spectrum-based methods can also predict resistance across hospitals, but require laboratory measurements after specimen processing and therefore operate later in the clinical workflow\cite{weis2022maldi}.

Five limitations recur across these studies. Most models were developed at one or two academic systems, and performance falls when they are applied at a different hospital or to a later period\cite{kim2022machine}. Probability calibration is assessed incompletely, and subgroup performance is rarely audited. Most also draw only on data recorded during the index encounter, omitting established risk factors that require additional assembly: prior antibiotic exposure is captured in structured medication orders but must be aggregated across a patient's earlier encounters, and neighborhood socioeconomic deprivation requires linkage to an external index. Both carry predictive information; exposure to four or more antibiotics in the preceding six months was associated with increased odds of multidrug-resistant community urine isolates (odds ratio, 6.81)\cite{malcolm2018risk}, and the Area Deprivation Index (ADI), a validated ranking of neighborhood socioeconomic disadvantage derived from census block-group measures of income, education, employment and housing quality\cite{kind2018making}, adds signal beyond clinical variables\cite{cooper2024socioeconomic,diaz2025integrating}. Infectious Diseases Society of America (IDSA) guidance recommends considering prior organisms, susceptibility results and recent antibiotic exposure, but does not prescribe a patient-specific method for integrating these signals during the empiric interval\cite{tamma2024idsa}.

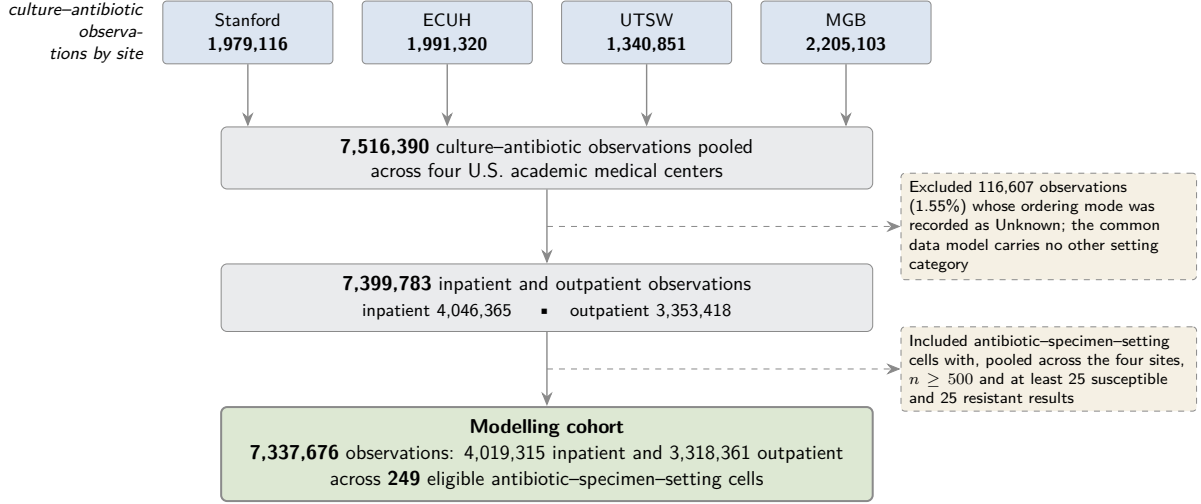
\begin{figure}[htbp]
\centering
\resizebox{\textwidth}{!}{%
\begin{tikzpicture}[
  font=\sffamily\small,
  >={Stealth[length=2.5mm,width=2mm]},
  sitebox/.style ={rounded corners=2pt, draw=cfrule, fill=cfSite, align=center,
                   text width=2.7cm, minimum height=1.1cm, inner sep=4pt,
                   font=\sffamily\footnotesize},
  mainbox/.style ={rounded corners=3pt, draw=cfrule, fill=cfMain, align=center,
                   text width=11cm, minimum height=1.0cm, inner sep=6pt},
  finalbox/.style={rounded corners=3pt, draw=cfrule, line width=1pt, fill=cfFinal,
                   align=center, text width=11cm, minimum height=1.2cm, inner sep=6pt},
  notebox/.style ={rounded corners=2pt, draw=cfrule, dashed, fill=cfNote, align=left,
                   text width=4.9cm, inner sep=4pt, font=\sffamily\scriptsize},
  arrow/.style   ={->, line width=0.7pt, draw=cfrule},
  dnote/.style   ={->, line width=0.5pt, draw=cfrule, dashed},
]
\node[sitebox] (s1) {Stanford\\[1pt]\textbf{1,979,116}};
\node[sitebox, right=0.5cm of s1] (s2) {ECUH\\[1pt]\textbf{1,991,320}};
\node[sitebox, right=0.5cm of s2] (s3) {UTSW\\[1pt]\textbf{1,340,851}};
\node[sitebox, right=0.5cm of s3] (s4) {MGB\\[1pt]\textbf{2,205,103}};
\node[font=\sffamily\footnotesize\itshape, anchor=east, align=right, text width=2.6cm]
  at ([xshift=-6pt]s1.west) {culture--antibiotic observations by site};
\coordinate (rowmid) at ($(s1)!0.5!(s4)$);
\node[mainbox] (total) at ([yshift=-2.2cm]rowmid)
  {\textbf{7,516,390} culture--antibiotic observations pooled across four
   U.S.\ academic medical centers};
\node[mainbox, below=1.3cm of total] (io)
  {\textbf{7,399,783} inpatient and outpatient observations\\[1pt]
   \footnotesize inpatient 4,046,365 \quad\textbullet\quad outpatient 3,353,418};
\node[finalbox, below=1.3cm of io] (final)
  {\textbf{Modelling cohort}\\[1pt]
   \textbf{7,337,676} observations: 4,019,315 inpatient and 3,318,361 outpatient\\[1pt]
   across \textbf{249} eligible antibiotic--specimen--setting cells};
\foreach \s in {s1,s2,s3,s4} \draw[arrow] (\s.south) -- (\s.south |- total.north);
\draw[arrow] (total.south) -- (io.north);
\draw[arrow] (io.south) -- (final.north);
\coordinate (exmid)  at ($(total.south)!0.5!(io.north)$);
\coordinate (incmid) at ($(io.south)!0.5!(final.north)$);
\node[notebox, anchor=west] (ex)  at ([xshift=6.2cm]exmid)
  {Excluded 116,607 observations (1.55\%) whose ordering mode was recorded as
   Unknown; the common data model carries no other setting category};
\node[notebox, anchor=west] (inc) at ([xshift=6.2cm]incmid)
  {Included antibiotic--specimen--setting cells with, pooled across the four
   sites, $n\geq500$ and at least 25 susceptible and 25 resistant results};
\draw[dnote] (exmid)  -- (ex.west);
\draw[dnote] (incmid) -- (inc.west);
\end{tikzpicture}%
}
\caption{\textbf{Study cohort and inclusion flow.} The four sites contributed 7,516,390 culture--antibiotic observations. Exclusion of 116,607 observations with unknown ordering mode left 7,399,783 inpatient and outpatient observations. Antibiotic--specimen--setting cells required at least 500 pooled observations, including at least 25 susceptible and 25 resistant results. The final modeling cohort comprised 7,337,676 observations (4,019,315 inpatient and 3,318,361 outpatient) across 249 eligible cells.}
\label{fig:1}
\end{figure}

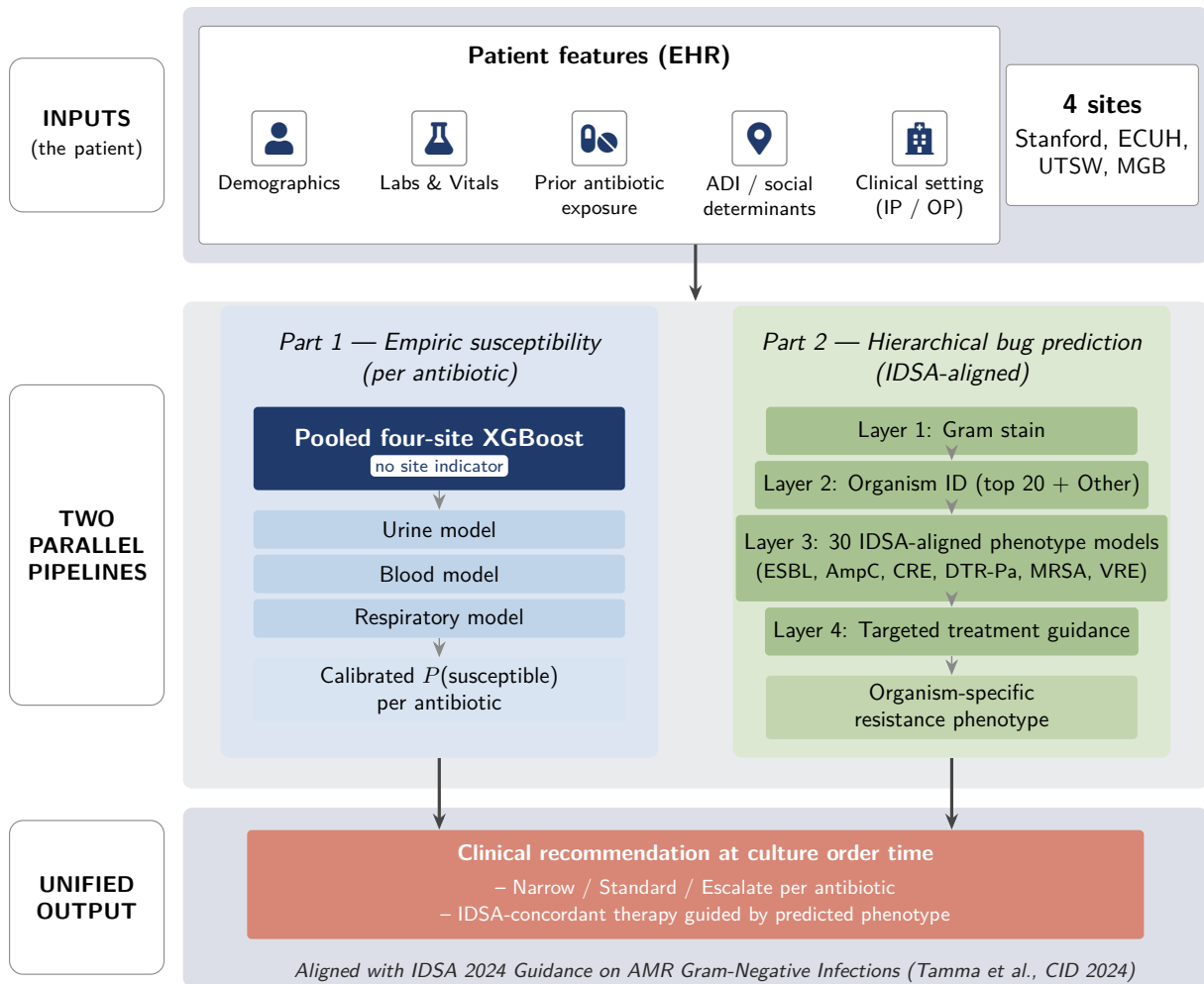
\begin{figure}[htbp]
\centering
\definecolor{tierBg}{HTML}{DCDFE7}
\definecolor{midBg}{HTML}{EAEBED}
\definecolor{partAbg}{HTML}{DCE5F1}
\definecolor{partBbg}{HTML}{DDE8D3}
\definecolor{xgblue}{HTML}{1F3A6B}
\definecolor{modelblue}{HTML}{BFD3E9}
\definecolor{caliblue}{HTML}{D8E4F1}
\definecolor{layergreen}{HTML}{A7C290}
\definecolor{layerlight}{HTML}{C2D5AE}
\definecolor{outred}{HTML}{D88675}
\definecolor{rule}{HTML}{8A8A8A}

\resizebox{\textwidth}{!}{%
\begin{tikzpicture}[
  font=\sffamily\small,
  >={Stealth[length=2.5mm,width=2mm]},
  card/.style    ={rounded corners=2pt, draw=rule, fill=white, align=center, inner sep=4pt},
  iconlabel/.style={align=center, font=\sffamily\footnotesize, text width=2.4cm},
  glyph/.style   ={rounded corners=2pt, draw=rule, line width=0.6pt,
                   minimum size=0.85cm, inner sep=0pt, fill=white,
                   text=xgblue, font=\Large},
  partheader/.style={font=\sffamily\itshape, align=center, text width=6.0cm},
  xgbox/.style   ={rounded corners=2pt, fill=xgblue, text=white, align=center,
                   minimum width=5.8cm, minimum height=1.3cm, font=\sffamily\bfseries},
  modelbox/.style={rounded corners=2pt, fill=modelblue, align=center,
                   minimum width=5.8cm, minimum height=0.6cm},
  calibox/.style ={rounded corners=2pt, fill=caliblue, align=center,
                   minimum width=5.8cm, minimum height=0.95cm},
  layerbox/.style={rounded corners=2pt, fill=layergreen, align=center,
                   minimum width=5.8cm, minimum height=0.75cm},
  phenbox/.style ={rounded corners=2pt, fill=layerlight, align=center,
                   minimum width=5.8cm, minimum height=0.95cm},
  outbox/.style  ={rounded corners=2pt, fill=outred, text=white, align=center,
                   minimum width=14.0cm, minimum height=1.7cm},
  arrow/.style   ={->, line width=0.6pt, rule},
  bigarrow/.style={->, line width=1.4pt, rule!50!black},
  panel/.style   ={rounded corners=4pt, inner sep=8pt},
  tierLabelCard/.style={
    rounded corners=6pt, fill=white, draw=rule, line width=0.5pt,
    align=center, text width=1.85cm,
    font=\sffamily\bfseries\small, inner sep=8pt,
  },
]

\begin{scope}[on background layer]
  \node[panel, fill=tierBg, minimum width=16cm, minimum height=4.0cm] (tier1)
    at (7, 11.4) {};
  \node[panel, fill=midBg,  minimum width=16cm, minimum height=7.6cm] (tier2)
    at (7,  5.0) {};
  \node[panel, fill=tierBg, minimum width=16cm, minimum height=2.8cm] (tier3)
    at (7, -0.5) {};
\end{scope}

\node[card, minimum width=12.5cm, minimum height=3.4cm] (inputbox) at (5.5,11.4) {};
\node[font=\sffamily\bfseries, anchor=north] at ([yshift=-5pt]inputbox.north)
  {Patient features (EHR)};
\node[glyph] (g1) at ([xshift=-5.0cm,yshift=-0.05cm]inputbox.center) {\faUser};
\node[iconlabel, anchor=north] at (g1.south) {Demographics};
\node[glyph] (g2) at ([xshift=-2.5cm,yshift=-0.05cm]inputbox.center) {\faFlask};
\node[iconlabel, anchor=north] at (g2.south) {Labs \& Vitals};
\node[glyph] (g3) at ([xshift=0.0cm,yshift=-0.05cm]inputbox.center)  {\faPills};
\node[iconlabel, anchor=north] at (g3.south) {Prior antibiotic\\ exposure};
\node[glyph] (g4) at ([xshift=2.5cm,yshift=-0.05cm]inputbox.center)  {\faMapMarker*};
\node[iconlabel, anchor=north] at (g4.south) {ADI / social\\ determinants};
\node[glyph] (g5) at ([xshift=5.0cm,yshift=-0.05cm]inputbox.center)  {\faHospital};
\node[iconlabel, anchor=north] at (g5.south) {Clinical setting\\ (IP / OP)};

\node[card, minimum width=2.8cm, minimum height=2.2cm] (sites) at (13.35,11.4)
  {\bfseries\large 4 sites\\[3pt]\normalsize\mdseries Stanford, ECUH,\\ UTSW, MGB};

\node[partheader] (h1) at (3,7.9)
  {Part 1 --- Empiric susceptibility\\ (per antibiotic)};
\node[xgbox, below=0.18cm of h1] (xgb) {};
\node[anchor=north, text=white, font=\sffamily\bfseries]
  at ([yshift=-6pt]xgb.north) {Pooled four-site XGBoost};
\node[rounded corners=2pt, fill=white, text=xgblue,
      font=\sffamily\scriptsize, inner sep=2pt, anchor=south]
  at ([yshift=6pt]xgb.south) {no site indicator};
\node[modelbox, below=0.30cm of xgb]   (urine) {Urine model};
\node[modelbox, below=0.08cm of urine](blood) {Blood model};
\node[modelbox, below=0.08cm of blood](resp)  {Respiratory model};
\node[calibox,  below=0.30cm of resp]  (calib)
  {Calibrated $P$(susceptible)\\ per antibiotic};

\node[partheader] (h2) at (11,7.9)
  {Part 2 --- Hierarchical bug prediction\\ (IDSA-aligned)};
\node[layerbox, below=0.18cm of h2]   (L1) {Layer 1: Gram stain};
\node[layerbox, below=0.08cm of L1]   (L2) {Layer 2: Organism ID (top 20 + Other)};
\node[layerbox, below=0.08cm of L2, minimum height=1.35cm] (L3)
  {Layer 3: 30 IDSA-aligned phenotype models\\[1pt]
   \footnotesize(ESBL, AmpC, CRE, DTR-Pa, MRSA, VRE)};
\node[layerbox, below=0.08cm of L3]   (L4) {Layer 4: Targeted treatment guidance};
\node[phenbox,  below=0.30cm of L4]   (phen)
  {Organism-specific\\ resistance phenotype};

\begin{scope}[on background layer]
  \node[panel, fill=partAbg,
        fit=(h1)(xgb)(calib)(phen.south -| calib)]  (p1panel) {};
  \node[panel, fill=partBbg, fit=(h2)(L1)(phen)]    (p2panel) {};
\end{scope}

\node[outbox] (out) at (7, -0.3)
  {\textbf{Clinical recommendation at culture order time}\\[3pt]
   \footnotesize -- Narrow / Standard / Escalate per antibiotic\\
   \footnotesize -- IDSA-concordant therapy guided by predicted phenotype};
\node[font=\sffamily\itshape\footnotesize, anchor=north east,
      text=rule!30!black] (idsanote)
  at ([yshift=-6pt]out.south east)
  {Aligned with IDSA 2024 Guidance on AMR Gram-Negative Infections
   (Tamma et al., CID 2024)};

\node[tierLabelCard, anchor=east, minimum height=2.4cm] (lab1)
  at ([xshift=-8pt]tier1.west) {INPUTS\\[1pt]\footnotesize\mdseries (the patient)};
\node[tierLabelCard, anchor=east, minimum height=5.0cm] (lab2)
  at ([xshift=-8pt]tier2.west) {TWO\\ PARALLEL\\ PIPELINES};
\node[tierLabelCard, anchor=east, minimum height=2.4cm] (lab3)
  at ([xshift=-8pt]tier3.west) {UNIFIED\\ OUTPUT};

\draw[bigarrow] (inputbox.south -| tier2) -- (tier2.north);
\draw[arrow]    (xgb.south)      -- (urine.north);
\foreach \a/\b in {L1/L2, L2/L3, L3/L4, L4/phen, resp/calib}
  \draw[arrow] (\a.south) -- (\b.north);
\draw[bigarrow] (p1panel.south) -- (p1panel.south |- out.north);
\draw[bigarrow] (p2panel.south) -- (p2panel.south |- out.north);

\end{tikzpicture}%
}
\caption{\textbf{ARMD-Predict framework.} A shared EHR feature set feeds two research pipelines. The empiric-susceptibility component estimates the probability that a recovered isolate is susceptible to each candidate antibiotic; released pooled models omit site identity, and the public interface uses the short-form feature set. Calibration transforms were fitted on validation data, but transform selection reused test-set ECE. The organism-and-resistance component classifies Gram stain, ranks organism labels and estimates organism-specific resistance phenotypes. Phenotype AUROCs were evaluated conditional on the observed organism and do not measure the end-to-end hierarchy. Phenotypic proxies align, where applicable, with 2024 IDSA guidance; MRSA and VRE are Gram-positive complements. Models were developed using data from Stanford Health Care, ECUH, UTSW and MGB.}
\label{fig:2}
\end{figure}

We developed ARMD-Predict using 7.5 million culture--antibiotic observations from four U.S. academic medical centers (Fig.~\ref{fig:1}); Stanford Health Care, East Carolina University Health (ECUH), UT Southwestern Medical Center (UTSW) and Mass General Brigham (MGB). An empiric-susceptibility predictor estimates whether a candidate antibiotic will be active against a recovered isolate. A complementary hierarchical model predicts the infecting organism and clinically relevant resistance phenotypes aligned with 2024 IDSA guidance (Fig.~\ref{fig:2}). We evaluated discrimination, transportability, calibration, projected prescribing utility and subgroup performance, and released both components as publicly accessible research applications for external technical evaluation.

\section{Methods}\label{sec:methods}

\subsection{Ethical statement}\label{sec:ethical-statement}

This study used retrospectively collected, de-identified data and was reviewed and approved by the Stanford University Institutional Review Board (IRB \#70466) with a waiver of informed consent.

\subsection{Study sites and cohorts}\label{sec:study-sites-and-cohorts}

We assembled microbiology data from four U.S. academic medical centers: Stanford Health Care (Stanford, California), ECUH (Greenville, North Carolina), UTSW (Dallas, Texas) and MGB (Boston, Massachusetts)\cite{nateghi2025armd,cooper2025armdecuh,cooper2026armdutsw,wei2025armdmgb}. Site-specific culture dates were October 2007 to January 2024 at Stanford, November 2014 to February 2025 at ECUH and April 2006 to February 2025 at UTSW. MGB applied patient-specific date shifts that preserved within-patient intervals but precluded a calendar window. De-identified extracts linked culture orders and antimicrobial susceptibility results to demographics, care setting, prior antibiotics, procedures, organisms, resistance events, comorbidities, ADI and, where available, laboratory values and vital signs. Fig.~\ref{fig:1} summarizes cohort selection and Extended Data Table~\ref{edtab:1} summarizes the cohorts.

\subsection{Susceptibility labels and IDSA-aligned phenotypes}\label{sec:susceptibility-labels-and-idsa-aligned-phenotypes}

Susceptibility was binarized at the antibiotic--organism level: clinical laboratory susceptible (S) was coded \(y = 1\); resistant (R) and intermediate (I) were coded \(y = 0\). Grouping intermediate with resistant is a deliberate and conservative labeling policy rather than a convenience: an intermediate result does not license empiric use of the agent at standard dosing, so for the empiric question this framework addresses it is treated as a failure of coverage.

Labels derive from two sources. The majority are measured, in the sense that the clinical microbiology laboratory reported a result for that organism--antibiotic pair. The remainder are laboratory-inferred: susceptibility that was not tested directly for that isolate but that follows from the reported panel together with the organism's intrinsic or predictable resistance, propagated by a site-supplied implied-susceptibility rule table. The inferred share is not negligible and varies by site, from 7.1\% of observations at ECUH to 21.4\% at Stanford (14.5\% at UTSW, 17.7\% at MGB). Inferred labels are never allowed to overwrite a measured result; where a measured and an inferred label disagree for the same organism--antibiotic pair the measured label is used, and where two measured results disagree the more resistant one is retained.

When a culture grew more than one organism, each recovered organism contributes its own susceptibility result: the analysis cohort is organism-level, carrying one row per organism--antibiotic pair, and polymicrobial orders are not collapsed to a single row per order. Across the four sites, 317,798 order--antibiotic pairs (650,691 rows, 8.7\% of the cohort) involve more than one recovered organism, and 78,394 of those pairs carry discordant susceptibility across the organisms recovered. Because the empiric question is asked of the order rather than of the isolate, an agent that covers one recovered organism but not another is counted here as susceptible for the covered isolate and resistant for the uncovered one, rather than as a single failure of coverage for the order; the consequences of that choice are taken up in the Discussion.

Antibiotic names were normalized across sites to a canonical vocabulary (for example, \emph{pip/tazo} was mapped to piperacillin--tazobactam). For the organism-specific resistance-phenotyping stage of the organism-and-resistance predictor (described below), organism-specific resistance phenotypes were defined to align with the 2024 IDSA guidance on antimicrobial-resistant Gram-negative infections\cite{tamma2024idsa}: ceftriaxone non-susceptibility as a proxy for ESBL production (section 1); cefepime/ceftriaxone non-susceptibility in \emph{Enterobacter cloacae} complex, \emph{Klebsiella aerogenes}, and \emph{Citrobacter freundii} as a proxy for AmpC \(\beta\)-lactamase derepression (section 2); ertapenem non-susceptibility as a proxy for carbapenem-resistant Enterobacterales (CRE; section 3); and joint resistance to multiple anti-pseudomonal agents as a proxy for difficult-to-treat \emph{P.~aeruginosa} (section 4), with MRSA (oxacillin non-susceptibility) and VRE (vancomycin non-susceptibility) included as Gram-positive complements.

\subsection{Data dictionary for empiric prescribing}\label{sec:data-dictionary-for-empiric-prescribing}

We defined empiric appropriate therapy as an antibiotic to which the recovered isolate was susceptible. Narrowing denoted selection of a narrower-spectrum agent that retained isolate coverage, and escalation denoted selection of a broader-spectrum or alternative agent. IDSA-aligned denoted consistency with the phenotype-based treatment recommendations in the 2024 IDSA guidance. These are microbiological definitions and do not establish patient-level treatment effects.

\subsection{Feature construction and feature-set policy}\label{sec:feature-construction-and-feature-set-policy}

We constructed 13 candidate feature blocks from a harmonized common data model: demographics; care setting; comorbidities; prior antibiotics by class and subtype; prior procedures; nursing-home history; prior organisms; prior resistance events; same-antibiotic resistance within 365 days; ADI; and, where available, laboratory values and vital signs. Candidate blocks were considered when available at three or more sites; the final pooled feature matrix applied the all-site restrictions described below.

Most blocks are event-level, with one row per recorded event and the interval to the culture order in whole days. Each block was reduced to one row per culture order by one of four procedures. Blocks with no day offset and one row per order already (demographics, care setting, ADI, laboratory values) were carried through unchanged. Comorbidities were expanded to multi-hot indicators, so that every component recorded for an order contributes its own binary column, together with a count of distinct components. Nursing-home contact was reduced to presence alone, because its day offsets were dominated by same-day records. The remaining history blocks (prior antibiotic exposure by agent, class and subtype, prior invasive procedures, prior infecting organisms and prior resistance events) were summarized as a count of qualifying events, the number of days since the most recent event, an indicator for any prior event, and indicators for any event within 30, 90 and 365 days.

Laboratory values are available at all four sites. Vital signs are available at Stanford, ECUH, and MGB but not at UTSW. Because the pooled training matrix is the intersection of the columns the contributing sites share, the vital-sign block is dropped rather than filled, and none of the 23 released models uses a vital sign. The consequences are taken up in the Discussion.

Two rules were applied to every history block. First, records with a day offset of zero were excluded. Whole-day offsets cannot establish that a same-day event preceded the order, so these records were removed at all sites, including MGB, where timestamps are exact. The rule changed the data materially: before exclusion, every Stanford prior-medication record retained by first-row selection carried a zero offset, which described 53.8\% of orders as having received an antibiotic on the day of culture, and the median interval since the most recent antibiotic rose from 0 to 70 days once these records were removed. Second, where a block also contributed a descriptive value, such as the name of an antibiotic or a procedure, the value was taken from the most recent qualifying event rather than the first record listed. Extracts list same-day records first, so first-row selection systematically returned the record closest to the order.

Categorical values were one-hot encoded. Comorbidity components are the only block represented in full: for every other block the descriptive value describes the most recent qualifying event only, so a patient with several prior antibiotics is represented by the count, the recency and the identity of the most recent agent. The identity of prior infecting organisms was not retained in any form. Counts and durations of prior hospitalization were not available in the harmonized data model at all four sites and were therefore not included. Columns carrying shifted or raw timestamps were removed by an automated leakage check before training. Extended Data Table~\ref{edtab:5} lists the retained feature blocks and their column counts, and the complete column list is provided as a supplementary file.

\subsection{Missing data}\label{sec:missing-data}

Missing values were not otherwise imputed, and absence was encoded distinctly from zero. For time-since-event variables subject to monotonic constraints, absence was encoded as 99,999 days as described below. Qualifying same-day records were excluded as described above. Event counts and any-within-window flags were zero when no prior qualifying event was recorded. Other empty numeric cells were passed to XGBoost, which learns a default branch for missing values at each split. Missing categorical values collapsed to a single unknown level encoded alongside the observed levels.

For these monotonically constrained time-since-event variables, absence of a prior event was encoded as 99,999 days so that it lay at the end of the constrained axis rather than on XGBoost's default missing branch. The same transformation was applied during training, prediction, calibration, operating-point analysis, subgroup analysis and SHAP computation; model metadata recorded the affected columns and fill value.

Feature blocks absent at a contributing site were dropped rather than filled. The pooled matrix retained 610 of 677 candidate columns; excluded columns comprised site-limited vital-sign, laboratory, comorbidity and timing variables. Missingness in retained laboratory, ADI and prior-event variables was substantial and varied by care setting.

\subsection{Data splitting}\label{sec:data-splitting}

The 23 pooled models used a random, patient-grouped 60/20/20 training, validation and test split (seed 20260806), preventing a patient from contributing observations to more than one partition. Because the split was contemporaneous, it does not estimate future-period performance. In 21 matched site--antibiotic analyses, chronological splitting reduced AUROC by a mean of 0.0396 and median of 0.0411. The 24 Stanford-only models instead used patient-grouped chronological splits. The transportability strategy comparison used the same random patient-grouped partition and shares its temporal limitation. The organism and phenotype models used the distinct splits described below.

\subsection{Development and validation of the empiric susceptibility models}\label{sec:development-and-validation-of-the-empiric-susceptibility-models-1}

We trained one XGBoost empiric-susceptibility model per antibiotic, specimen and care-setting cell (500 estimators, maximum depth 6, learning rate 0.05, subsampling and column sampling 0.8, and early stopping on validation data). Monotonic constraints required predicted susceptibility not to increase with additional prior resistance or antibiotic exposure and not to decrease with time since those events. Cells required at least 500 pooled observations, including at least 25 susceptible and 25 resistant results; 249 of 445 candidate cells qualified. We report 23 pooled four-site models, each with full and short-form feature variants, and a separate family of 24 chronologically evaluated Stanford-only models. These families used different data and splits. Multi-site strategy analyses compared a Stanford-only baseline, pooled models with and without a site indicator, and GroupDRO with site as the group\cite{sagawa2020distributionally}.

The organism-and-resistance predictor classifies each specimen through three sequential stages: a hierarchical design that proceeds from Gram-stain classification, to organism identification, to organism-specific resistance phenotyping. The Gram-stain stage used binary XGBoost classifiers for Gram-positive vs.~Gram-negative status per specimen type; the organism-identification stage used multi-class XGBoost models over 21 classes, the 20 most frequently recovered organisms in that specimen and setting stratum plus an \emph{Other} class; and the resistance-phenotyping stage trained 30 organism-specific binary classifiers on the IDSA-aligned phenotypes described above. Two properties of the organism-identification stage should be read alongside its accuracy. The class list is stratum-specific rather than fixed, so an organism such as \emph{Enterococcus faecium} is resolved separately in the blood and inpatient urine strata but falls into \emph{Other} elsewhere. Organism labels were also taken as the clinical laboratories reported them and were not harmonized to species, so in some strata a single organism occupies two adjacent class labels, which lowers top-1 accuracy for those strata by construction.

\subsection{Transportability analysis across hospitals}\label{sec:transportability-analysis-across-hospitals}

Transportability analyses compared Stanford-only training, pooled four-site training with and without a site indicator, and GroupDRO with site as the group\cite{sagawa2020distributionally}. Only the site-aware pooled development model included hospital identity. Publicly released models use pooled data without a site feature.

Evaluation scope differed by training strategy. Models evaluated at a site that contributed training data were scored on that site's test subset, whereas models evaluated at a site excluded from training were scored on its full cohort. Stanford-only and pooled models at MGB were therefore evaluated on different row sets (median 18,869 versus 3,729 cultures across 23 cells), and that comparison is not patient-matched. Comparisons among pooled strategies used the same test rows.

We also performed four leave-one-site-out runs using the same patient-grouped partition. Only the MGB held-out results were retained, with each strategy evaluated on identical cultures. GroupDRO was not repeated in this design. Selection among pooled variants used mean AUROC at MGB and therefore was not independent of evaluation at that site.

\subsection{Probability calibration}\label{sec:probability-calibration}

Calibration was evaluated separately for the chronologically split Stanford-only models and the contemporaneously split pooled models. Candidate isotonic and beta-calibration transforms were fitted on validation data\cite{kull2017beta}. In the current analysis, however, the transform or untransformed output was selected by lowest test-set ECE and that test-set ECE was then reported. Because transform selection and evaluation reused the same test set, the reported calibration estimates are optimistic and do not constitute independent validation. A single global curve was fitted per model; no site-specific calibration was performed. Both transforms were monotonic and therefore did not alter AUROC. Model-family calibration results are reported in the Results.

\subsection{Local-prevalence re-anchoring}\label{sec:local-prevalence-re-anchoring}

Recalibration aligns probabilities to the pooled training prevalence \(\pi_{train}\), which is itself baked into every calibrated output. A site whose baseline resistance differs from \(\pi_{train}\) will therefore see systematically miscalibrated probabilities even when the model ranks patients correctly. To allow deployment at outside hospitals \emph{without retraining}, the web application optionally re-anchors each prediction to the user's local prevalence \(\pi_{local}\) (e.g.~from their antibiogram) via a prior-probability-shift correction\cite{saerens2002adjusting}. Working in odds, we back out a per-patient likelihood ratio from the calibrated output \(p_{model} = P(R \mid x)\) and re-apply it to the local prior:

\begin{equation*}
\operatorname{logit}(p_{\mathrm{local}}) = \operatorname{logit}(p_{\mathrm{model}}) - \operatorname{logit}(\pi_{\mathrm{train}}) + \operatorname{logit}(\pi_{\mathrm{local}}).
\end{equation*}

The correction assumes prior-probability shift: the class-conditional feature distributions \(P(x \mid R)\) and \(P(x \mid S)\) (equivalently the per-patient likelihood ratio) are stable across sites, and only the marginal prevalence \(P(R)\) differs. Because the transform is monotonic in \(p_{model}\), it shifts probabilities without altering rank-order, leaving AUROC unchanged by construction.

We evaluated the correction by leave-one-site-out retraining. For each site in turn, an XGBoost model was trained on the remaining three sites (omitting the site-prevalence feature, so the shift is a purely external adjustment), isotonic-recalibrated on a held-out slice of the training sites, and applied to the unseen site; predictions carrying the three-site training prior were compared with predictions re-anchored to the held-out site's marginal prevalence by ECE, Brier score, calibration slope and intercept, and AUROC. A controlled prevalence sweep on the out-of-sample predictions, resampled to target prevalence while preserving the within-class score distributions, quantified the correction under larger prevalence gaps than the four academic sites span.

\subsection{Operating thresholds for stewardship and safety}\label{sec:operating-thresholds-for-stewardship-and-safety}

For each empiric-susceptibility model, we explored two operating thresholds on the held-out test observations. The narrowing threshold was the lowest predicted-susceptibility cutoff with positive predictive value of at least 95\% and at least 20 predicted-susceptible observations; yield was the proportion of susceptible isolate observations above this threshold. The escalation threshold was the highest cutoff at which no more than 20\% of susceptible observations were flagged; yield was the proportion of resistant observations flagged. Models that did not attain 95\% positive predictive value had no narrowing operating point. Because thresholds and yields were derived from the same test observations, these estimates require validation in independent data. The unit of analysis was the culture--organism--antibiotic observation, not the patient or culture order.

\subsection{Subgroup analysis}\label{sec:subgroup-analysis}

We prespecified subgroup analyses of test-set AUROC by ADI quintile, sex and age band across the 23 reported models. We calculated 95\% CIs using 1,000 stratified bootstrap replicates. These analyses evaluated discrimination only, not calibration or operating-threshold performance, and subgroup similarity was not interpreted as evidence of equitable clinical effects\cite{pfohl2024toolbox}.

\subsection{Organism and resistance-phenotype pipeline}\label{sec:organism-and-resistance-phenotype-pipeline}

We fitted three additional model layers. Organism classifiers predicted the 20 most frequent organism labels in each specimen and setting stratum plus an Other class; 12 specimen--setting configurations were specified, of which ten completed variants are reported in the Results. Organism-informed susceptibility models appended 40 organism indicators to the pre-result feature matrix and included 50 antibiotic-specific models. Thirty models predicted organism-specific resistance phenotypes. The latter two layers used the reported organism label and were not evaluated as an end-to-end cascade from predicted organism to phenotype.

Resistance-phenotype targets combined a case-insensitive organism-name pattern with an antibiotic result and treated resistant or intermediate results as positive for resistance. Targets were phenotypic proxies aligned with the 2024 IDSA guidance where applicable\cite{tamma2024idsa}; MRSA and VRE were included as Gram-positive complements. Of 34 configured targets, 30 met training thresholds of at least 100 observations and at least 20 per class. Models used XGBoost (seed 42) with early stopping on validation data and feature blocks available at three or more sites.

Organism and phenotype models used calendar cutoffs at Stanford, ECUH and UTSW (training before 1 January 2020, validation through 1 January 2022 and test thereafter) and a deterministic 70/15/15 patient-hash split at MGB, where dates were shifted. Calendar splitting was not patient-grouped, so patients with cultures spanning a cutoff could contribute to more than one partition. Rows without usable timestamps inherited a patient assignment or, when unavailable, a patient hash. Site-specific partitions were pooled for training.

Specimen type did not enter Layers 2 and 3 as a feature: both were trained on the unfiltered combined frame, and neither feature list contains a culture-type column. Care setting entered through four ward indicators (emergency department, intensive care, inpatient, outpatient).

\subsection{Model interpretation}\label{sec:model-interpretation}

Model interpretability was assessed with TreeExplainer SHAP values\cite{lundberg2017unified} computed on held-out test data: for each released empiric model, a seeded subsample of up to 2,000 records from its recorded patient-grouped test slice; for each resistance-phenotype model, the organism pipeline's pooled test split restricted to that model's own isolates; and for the Gram-stain stage, a seeded subsample of the same pooled test split. Features were ranked by mean absolute SHAP value, with summary panels in Extended Data Fig.~\ref{edfig:6}.

\subsection{Resistance benchmarking}\label{sec:resistance-benchmarking}

Site-level resistance rates for the IDSA-aligned phenotypes were compared against published national surveillance estimates from the Centers for Disease Control and Prevention (CDC) Antibiotic Resistance Threats Report (2019)~(ref.~\citenum{cdc2019arthreats}) and the National Healthcare Safety Network (NHSN), using the same organism groupings and time windows where possible.

\subsection{Research web applications}\label{sec:research-web-applications}

The public Streamlit application at \url{https://armd-predict.stanford.edu} serves 23 antibiotic, culture and care-setting cells. Each cell has a full-feature variant and a short-form variant restricted to variables intended to be available at culture ordering. The public empiric page uses the short-form models; full-feature models remain available only in the internal build. The interface is a research prototype and has not been prospectively validated or integrated into an EHR.

All 46 model variants used the monotonic constraints described above. The same 99,999-day encoding for no prior event was applied at training and prediction. Features outside the prior-resistance, prior-infection and prior-antibiotic-exposure families were unconstrained.

Calibration labels in the interface indicate whether a model serves a transformed or untransformed probability. Because the current transform-selection procedure reused the test set, these labels should not be interpreted as independent calibration validation.

The application displays model-specific probability thresholds and offers experimental local-prevalence re-anchoring. Seventeen models attained the prespecified 95\% positive predictive value for narrowing in the study analysis. For the other six, the current implementation records a capped 0.10 cutoff and labels it untested; no narrowing operating point is reported for those models here. The current interface is therefore a research prototype and should not be used for clinical decision-making.

\subsection{Statistical analysis}\label{sec:statistical-analysis}

We treated susceptibility as the positive class. Per-model analyses included AUROC, area under the precision--recall curve (AUPRC), specificity at sensitivities of 0.90 and 0.80, Brier score, calibration intercept and slope, and ECE across ten equal-width probability bins. Positive predictive value was used only for the exploratory narrowing threshold. Evaluation sets differed among the pooled empiric models, chronological Stanford-only models, organism and phenotype models, and leave-one-site-out analyses as described in the corresponding subsections.

We summarized performance across models using medians and interquartile ranges because test-set size varied across cells. The subgroup audit used 1,000 class-stratified bootstrap replicates and percentile 95\% CIs; levels with fewer than 30 observations or fewer than five observations in either class were not scored. No hypothesis tests or multiplicity corrections were applied. Analyses used Python 3.11.2, XGBoost 3.1.3, scikit-learn 1.8.0, SHAP 0.50.0, NumPy 2.3.5 and pandas 2.3.3.

\section{Results}\label{sec:results}

\subsection{Baseline characteristics of the study cohorts}\label{sec:baseline-characteristics-of-the-study-cohorts}

The cohort included 7,516,390 culture--antibiotic observations from adults with urine, blood or respiratory cultures at four U.S. academic medical centers (Fig.~\ref{fig:1}). Stanford contributed 1,979,116 observations collected from October 2007 to January 2024, ECUH 1,991,320 from November 2014 to February 2025 and UTSW 1,340,851 from April 2006 to February 2025. MGB contributed 2,205,103 observations; patient-specific date shifts precluded a calendar window. After excluding 116,607 observations for which the care setting was not recorded as inpatient or outpatient, 7,399,783 observations remained. We evaluated antibiotic--specimen--setting cells containing at least 500 observations, including at least 25 susceptible and 25 resistant results. Of 445 candidate cells, 249 met these criteria. The final modeling cohort contained 7,337,676 observations, comprising 4,019,315 inpatient and 3,318,361 outpatient observations (Fig.~\ref{fig:1}). Cohort characteristics are reported in Extended Data Table~\ref{edtab:1}. Resistance prevalence differed from national surveillance estimates: vancomycin-resistant \emph{E. faecium} prevalence was 54.3\% compared with approximately 30\% nationally, whereas MRSA and fluoroquinolone-resistant \emph{E. coli} prevalence were lower (34.0\% versus approximately 43\%, and 24.5\% versus approximately 28\%, respectively; Extended Data Table~\ref{edtab:2}).

\subsection{Development and validation of the empiric susceptibility models}\label{sec:development-and-validation-of-the-empiric-susceptibility-models}

Across the 23 pooled empiric-susceptibility models, median held-out AUROC was 0.77 (interquartile range, 0.69--0.82) for inpatient urine cultures and 0.70 (0.66--0.76) for inpatient blood cultures; the single inpatient respiratory model reached 0.82 (Table~\ref{tab:1}). Median AUROC was 0.71 (0.67--0.74) for outpatient urine cultures and 0.82 for the single outpatient respiratory model. Across the 23 reported cells, pooled test partitions contained 375,501 observations. Across 87 reportable site-specific cells, hospital-specific median AUROC ranged from 0.691 to 0.715 (Table~\ref{tab:1}). Performance varied more by antibiotic than by site. Ampicillin was the weakest inpatient urine model (median AUROC, 0.63). In the full cohort, this analysis included 173,677 results, of which 46.0\% were susceptible; Table~\ref{tab:1} separately reports the pooled test partition. Ampicillin susceptibility is determined largely by which organism is present: \emph{P. aeruginosa} and \emph{K. pneumoniae} are intrinsically resistant (0.0\% and 0.1\% susceptible), whereas \emph{E. faecalis} is nearly always susceptible (99.6\%). Because the organism is not yet known when empiric therapy is chosen, the model cannot use the strongest single predictor of the result, which limits the discrimination achievable for antibiotics of this kind.

\begin{table}[htbp]
\captionsetup{position=top}
\caption{\textbf{Held-out performance of 23 pooled empiric-susceptibility models.} Models were trained on four-site data without a site-identity feature using a random patient-grouped 60/20/20 split (seed 20260806). The positive class was susceptibility. AUROC and AUPRC are reported for susceptibility; Spec@90\% is the proportion of resistant observations identified at a threshold with 90\% sensitivity for susceptible observations. The site range recalculates AUROC within each hospital's test rows; cells with fewer than 25 observations in either class were omitted. Daggers denote ranges from which one or more such site cells were excluded (five cells in total). Across 87 reportable site-specific cells, median AUROC was 0.710: 0.691 at Stanford, 0.715 at ECUH, 0.706 at UTSW and 0.712 at MGB.}\label{tab:1}
\centering\footnotesize
\begin{tabular}{@{}lrrrrrr@{}}
\toprule
\textbf{Antibiotic} & \textbf{Test n} & \textbf{Susc. (\%)} & \textbf{AUROC} & \textbf{AUPRC} & \textbf{Spec@90\%} & \textbf{Per-site AUROC} \\
\midrule
\addlinespace[3pt]
\multicolumn{7}{@{}l}{\emph{Urine, inpatient}} \\
Ampicillin & 32,451 & 44.2 & 0.629 & 0.538 & 0.262 & 0.61--0.64 \\
Cefepime & 22,454 & 94.6 & 0.807 & 0.983 & 0.514 & 0.78--0.81 \\
Ceftriaxone & 27,960 & 86.5 & 0.773 & 0.948 & 0.448 & 0.75--0.77 \\
Gentamicin & 33,363 & 89.4 & 0.679 & 0.934 & 0.358 & 0.66--0.71 \\
Meropenem & 19,996 & 98.8 & 0.831 & 0.997 & 0.535 & 0.79--0.86 \\
Piperacillin--tazobactam & 29,190 & 95.7 & 0.699 & 0.978 & 0.307 & 0.63--0.71 \\
Vancomycin & 5,111 & 93.5 & 0.825 & 0.983 & 0.541 & 0.79--0.82 \\
\emph{Median (7 models)} &  &  & \emph{0.773} & \emph{0.978} & \emph{0.448} &  \\
\addlinespace[3pt]
\multicolumn{7}{@{}l}{\emph{Urine, outpatient}} \\
Amoxicillin--clavulanate & 16,919 & 82.0 & 0.638 & 0.876 & 0.229 & 0.58--0.65 \\
Cefazolin & 24,728 & 80.0 & 0.708 & 0.885 & 0.373 & 0.66--0.74 \\
Cefpodoxime & 17,106 & 93.0 & 0.733 & 0.966 & 0.406 & 0.70--0.75 \\
Ciprofloxacin & 26,668 & 79.7 & 0.743 & 0.897 & 0.477 & 0.72--0.75 \\
Levofloxacin & 25,578 & 81.1 & 0.747 & 0.908 & 0.465 & 0.71--0.75 \\
Nitrofurantoin & 29,600 & 81.6 & 0.662 & 0.882 & 0.303 & 0.63--0.67 \\
TMP-SMX & 27,864 & 77.1 & 0.673 & 0.856 & 0.317 & 0.63--0.70 \\
\emph{Median (7 models)} &  &  & \emph{0.708} & \emph{0.885} & \emph{0.373} &  \\
\addlinespace[3pt]
\multicolumn{7}{@{}l}{\emph{Blood, inpatient}} \\
Ampicillin & 4,704 & 33.4 & 0.637 & 0.449 & 0.215 & 0.54--0.66 \\
Cefepime & 3,432 & 92.9 & 0.704 & 0.963 & 0.354 & 0.62--0.75 \\
Ceftriaxone & 4,976 & 80.6 & 0.703 & 0.897 & 0.313 & 0.67--0.73 \\
Gentamicin & 6,489 & 90.6 & 0.656 & 0.939 & 0.302 & 0.63--0.68 \\
Meropenem & 4,116 & 94.3 & 0.835 & 0.986 & 0.575 & 0.74--0.82\textsuperscript{\textdagger{}} \\
Piperacillin--tazobactam & 3,543 & 94.3 & 0.658 & 0.964 & 0.272 & 0.65--0.68 \\
Vancomycin & 4,518 & 97.6 & 0.825 & 0.994 & 0.523 & 0.80--0.82\textsuperscript{\textdagger{}} \\
\emph{Median (7 models)} &  &  & \emph{0.703} & \emph{0.963} & \emph{0.313} &  \\
\addlinespace[3pt]
\multicolumn{7}{@{}l}{\emph{Respiratory, inpatient}} \\
Ceftriaxone & 1,702 & 40.6 & 0.825 & 0.735 & 0.558 & 0.70--0.85 \\
\addlinespace[3pt]
\multicolumn{7}{@{}l}{\emph{Respiratory, outpatient}} \\
Levofloxacin & 3,033 & 54.5 & 0.822 & 0.820 & 0.555 & 0.79--0.84\textsuperscript{\textdagger{}} \\
\textbf{All 23 models} & \textbf{375,501} &  & \textbf{0.708} & \textbf{0.934} & \textbf{0.373} &  \\
\bottomrule
\end{tabular}
\end{table}

\subsection{Transportability across hospitals}\label{sec:transportability-across-hospitals}

\begin{figure}[htbp]
\centering
\includegraphics[width=\textwidth]{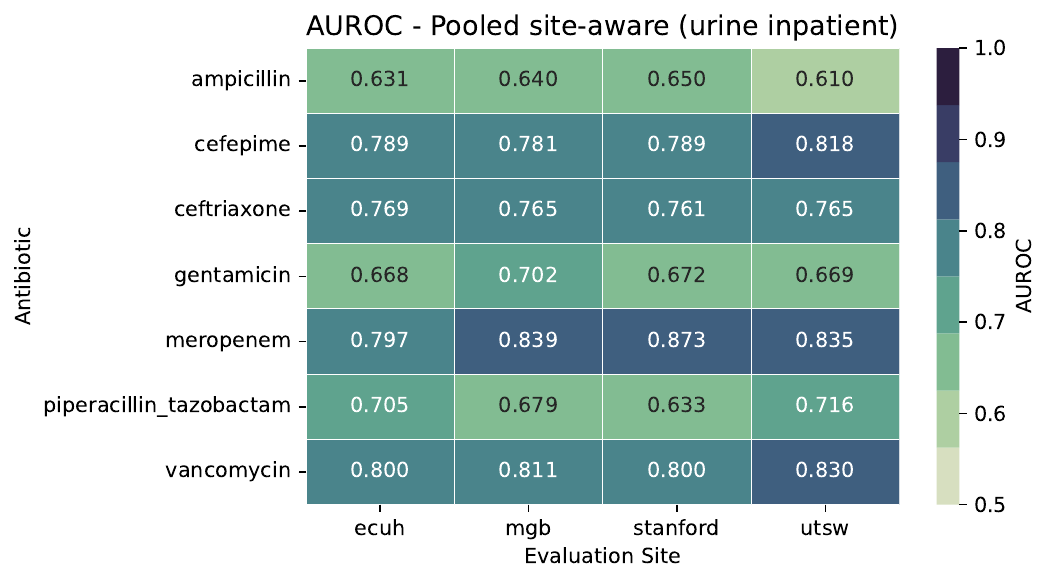}
\caption{\textbf{Discrimination of the site-aware pooled model across contributing sites.} The heatmap reports AUROC for seven inpatient urine antibiotics after training on data from all four hospitals with a one-hot site indicator. Each site contributed training data and was evaluated on its test subset from a random patient-grouped 60/20/20 split; this is not external validation at a hospital absent from training. Median AUROC across 28 antibiotic--site cells was 0.765 (interquartile range, 0.671--0.800; range, 0.610--0.873). Comparisons with the Stanford-only baseline in Extended Data Fig.~\ref{edfig:1} are descriptive because that model was evaluated on each external site's complete cohort rather than the pooled model's test subset. Released pooled models omit the site indicator.}
\label{fig:3}
\end{figure}

Stanford-only inpatient urine models transported unevenly across the three external hospitals. In the displayed panel, the largest AUROC decrease was from 0.838 to 0.738 for vancomycin at ECUH (Extended Data Fig.~\ref{edfig:1}). We therefore trained pooled models using data from all four hospitals. The site-aware development model achieved a median AUROC of 0.765 across 28 inpatient urine antibiotic--site cells (interquartile range, 0.671--0.800; range, 0.610--0.873; Fig.~\ref{fig:3}). Every hospital contributed training data and was evaluated on its own share of the same-period patient-grouped test partition; these results measure performance across represented hospitals, not transfer to a hospital absent from training. A site-agnostic pooled model performed nearly identically to the site-aware model (median difference, \textless0.001 AUROC), and released models omit site identity. Group distributionally robust optimization (GroupDRO) did not improve on pooling at MGB (median AUROC, 0.711 versus 0.720 across 24 cells).

Discrimination varied by specimen and setting. Median AUROC was 0.765 for inpatient urine (28 antibiotic--site cells; Fig. 3), 0.693 for outpatient urine (32 cells), 0.669 for inpatient blood (23 cells), 0.776 for inpatient respiratory (4 cells) and 0.797 for outpatient respiratory (6 cells) (Extended Data Fig. 7). The respiratory strata are supported by few cells because most antibiotic--site combinations did not meet inclusion thresholds, and their medians should be read accordingly.

\subsection{Combined urine and blood models for emergency workups}\label{sec:combined-urine-and-blood-models-for-emergency-workups}

A combined urine-and-blood model was intended for evaluations in which both cultures are ordered, removing the need to select a specimen-specific model at the time of empiric prescribing. Its performance was similar to that of the specimen-specific models. Median AUROC was 0.72 across 51 inpatient antibiotic comparisons, versus 0.73 for urine-specific and 0.68 for blood-specific models, and 0.71 across 47 outpatient comparisons, versus 0.70 and 0.63, respectively.

\subsection{Probability calibration and local-prevalence re-anchoring}\label{sec:probability-calibration-and-local-prevalence-re-anchoring}

Calibration assesses whether predicted probabilities agree with observed frequencies. Raw outputs from the 24 chronologically evaluated Stanford-only models were overconfident (median expected calibration error (ECE), 0.32; interquartile range, 0.23--0.43; median Brier score, 0.227). Isotonic recalibration reduced the reported median ECE to 0.014 and median Brier score to 0.115. The 23 pooled models were closer to calibrated before transformation (reported median ECE, 0.006). Candidate isotonic and beta transforms were fitted on validation data, but the final choice among a transform and the untransformed model was made using test-set ECE; the same test set was then used to report calibration performance. These post-selection estimates are therefore optimistic and require re-estimation after transform selection is confined to validation data. A single global curve was fitted per model, with no site-specific calibration. Monotonic recalibration did not change AUROC (Extended Data Fig.~\ref{edfig:2}).

The susceptibility application can also re-anchor predictions to a user-specified local resistance prevalence. In leave-one-site-out analyses, this prior-probability adjustment reduced ECE in 4 of the 5 site--model cells whose prevalence differed from training by at least 3 percentage points (mean ECE, 0.061 to 0.037), compared with improvement in 15 of all 20 cells (mean ECE, 0.044 to 0.033). In a controlled prevalence sweep, median ECE at a 40\% local resistance rate decreased from 0.20 to 0.06 after re-anchoring, without changing discrimination (Extended Data Fig.~\ref{edfig:3}). The larger gains in cells with a prevalence gap were consistent with the purpose of the adjustment. The retrospective site analysis used the realized resistance prevalence of the held-out site, rather than an independently sampled historical antibiogram, and should therefore be interpreted as an oracle-prevalence implementation.

\subsection{Clinical relevance of the operating thresholds}\label{sec:clinical-relevance-of-the-operating-thresholds}

\begin{figure}[htbp]
\centering
\includegraphics[width=0.92\textwidth]{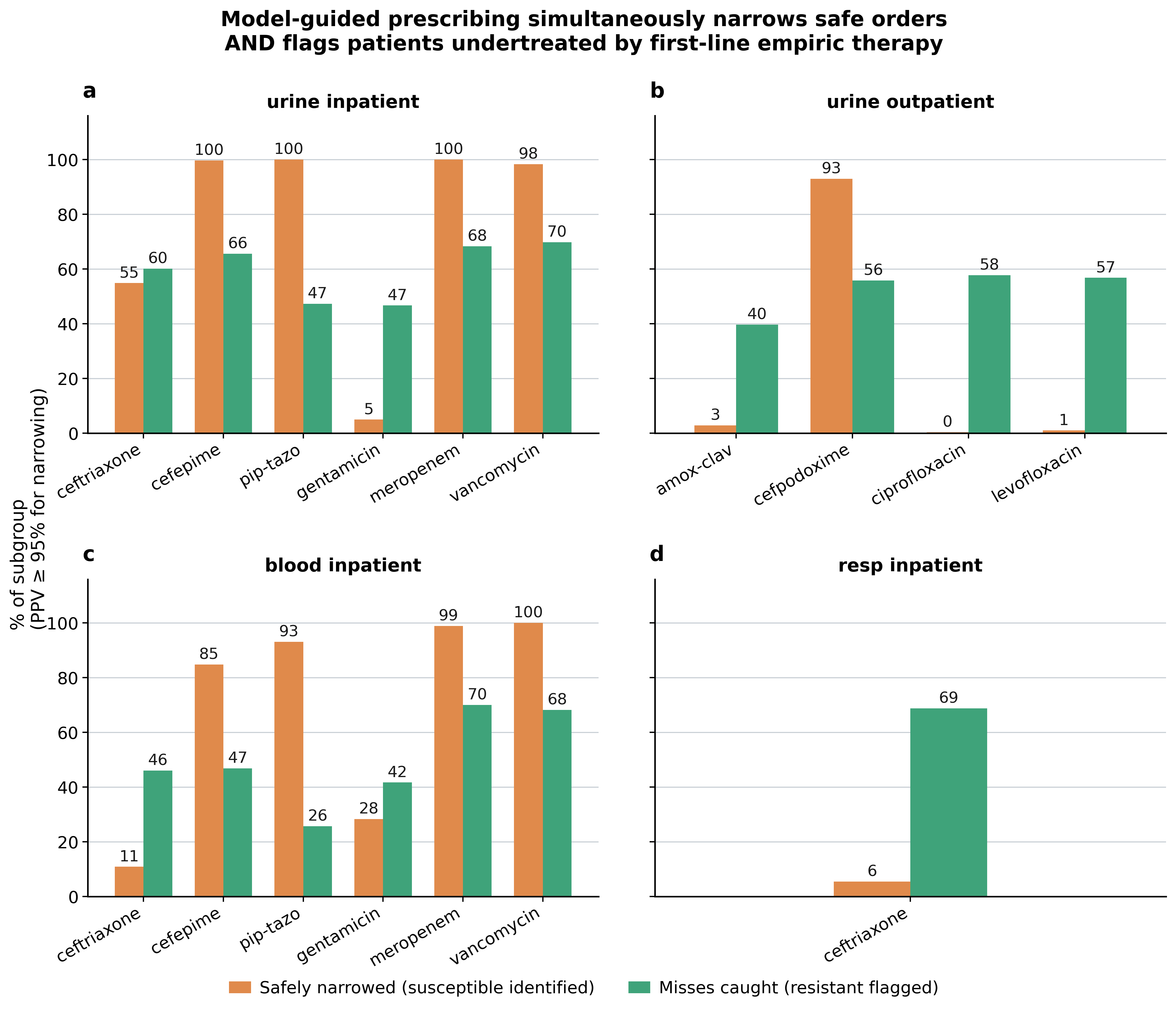}
\caption{\textbf{Exploratory narrowing and broader-coverage operating points.} Orange bars show the proportion of susceptible isolate observations identified for narrowing at a threshold with positive predictive value of at least 95\%. Green bars show the proportion of resistant isolate observations flagged for escalation or an alternative while limiting the proportion of susceptible observations flagged to 20\%. Thresholds and yields were derived and evaluated on the same test rows. For inpatient urine cultures, 54.9\% of ceftriaxone-susceptible and 5.0\% of gentamicin-susceptible observations were identified for narrowing, while 60\% and 47\% of resistant observations were flagged. Corresponding blood-culture values were 11\% and 28\% for narrowing and 46\% and 42\% for flagging. These are retrospective isolate-level microbiological-coverage projections, not patient-safety or treatment-effect estimates. Operating envelopes are shown in Extended Data Figs.~\ref{edfig:4} and \ref{edfig:5}.}
\label{fig:4}
\end{figure}

We next conducted an exploratory two-threshold prescribing analysis on test observations (Fig.~\ref{fig:4}). A high threshold identified observations for narrowing when positive predictive value was at least 95\%. A low threshold flagged resistant observations for broader coverage while limiting the proportion of susceptible observations flagged to 20\%. Both thresholds and their yields were derived and evaluated on the same test rows, so the results define candidate operating points rather than independently validated clinical rules. Outcomes are reported for culture--organism--antibiotic observations, not patients or culture orders.

Both thresholds were feasible for 17 of 23 models. Across these models, 26--70\% of resistant observations were flagged for broader coverage, whereas the proportion of susceptible observations identified for narrowing ranged from 0.3\% to 100\%. For inpatient urine cultures, 54.9\% of ceftriaxone-susceptible isolate observations (13,273 of 24,176) were identified for narrowing, while 60.1\% of ceftriaxone-resistant observations (2,274 of 3,784) were flagged for broader coverage, out of 27,960 test observations.

Corresponding values for gentamicin were 5.0\% and 47\%. In blood cultures, the corresponding ceftriaxone and gentamicin values were 11\% and 28\% for narrowing and 46\% and 42\% for broader-coverage flagging, respectively. For inpatient urine cefepime, piperacillin--tazobactam and meropenem, at least 99.7\% of susceptible observations were identified for narrowing and 47--68\% of resistant observations were flagged. In blood cultures, the largest projected combined yield was observed for meropenem (98.9\% narrowed and 70\% flagged). Three cells had baseline susceptibility above 95\%, so narrowing all observations met the predictive-value target without model discrimination. Among feasible outpatient urine models, only 0.3--2.8\% of susceptible observations were identified for narrowing with ciprofloxacin, levofloxacin and amoxicillin--clavulanate, compared with 93.0\% for cefpodoxime, reflecting baseline susceptibility of 80--82\% for the former three and 93.0\% for cefpodoxime.

The six models without a narrowing operating point were ampicillin in inpatient blood and urine cultures; cefazolin, nitrofurantoin and trimethoprim--sulfamethoxazole (TMP-SMX) in outpatient urine cultures; and levofloxacin in outpatient respiratory cultures. The urine and blood operating envelopes are shown in Extended Data Figs.~\ref{edfig:4} and \ref{edfig:5}. These estimates assume adherence to model output and ignore competing clinical considerations; they are retrospective microbiological-coverage projections, not patient-level safety or treatment-effect estimates.

\subsection{Performance across demographic subgroups}\label{sec:performance-across-demographic-subgroups}

In the prespecified subgroup audit, the median within-model AUROC range was 0.02 by sex, 0.07 by age band and 0.05 by ADI quintile (Extended Data Table~\ref{edtab:3}). The largest ranges were 0.279 by age band for vancomycin in inpatient blood cultures and 0.129 by ADI for cefepime in inpatient blood cultures. Restricting analyses to subgroups with at least 1,000 cultures reduced the median age-band range from 0.07 to 0.02, indicating that small subgroups contributed to the largest observed differences. Performance was not consistently ordered by ADI. Because race and ethnicity were unavailable and sex labels were inferred at three sites, this audit covers only a limited set of demographic axes. Similar discrimination across measured subgroups does not establish equitable clinical effects and warrants prospective monitoring\cite{pfohl2024toolbox}.

\subsection{Organism and resistance-phenotype prediction}\label{sec:organism-and-resistance-phenotype-prediction}

The organism-and-resistance component used a three-stage hierarchy (Fig.~\ref{fig:2}). The Gram-stain stage achieved pooled held-out AUROC 0.72, with site-specific values of 0.69--0.78; this degree of discrimination can shift but not determine the Gram-positive versus Gram-negative pathway. Organism classifiers predicted the 20 most frequent labels in each specimen and setting stratum plus an Other class. Across ten completed variants, top-1 accuracy ranged from 25.1\% to 67.0\% and top-3 accuracy from 51.3\% to 83.3\%. Four variants retained duplicate labels for the same organism, which divided probability across labels. \emph{E. coli} was the most frequent organism (212,743 culture orders), but its class-level AUROC was 0.71 in the inpatient urine model, within a 0.58--0.98 range across 21 classes.

Thirty organism-specific models predicted resistance phenotypes aligned with 2024 IDSA guidance where applicable, with \emph{MRSA} and vancomycin-resistant enterococci (VRE) included as Gram-positive complements (Table~\ref{tab:2}; Extended Data Table~\ref{edtab:4}). Across the complete set, AUROC ranged from 0.519 to 0.834 and test-set size from 657 to 48,290 observations. These models were evaluated conditional on the observed organism label, rather than as an end-to-end cascade through the predicted Gram class and organism. Ceftriaxone non-susceptibility, used as an ESBL proxy, reached AUROC 0.759 in \emph{E. coli}, 0.793 in \emph{K. pneumoniae} and 0.719 in \emph{P. mirabilis}; \emph{E. coli} ciprofloxacin resistance reached 0.746. Carbapenem-resistance targets had prevalence below 1\% and are reported as sensitivity analyses. For \emph{P. aeruginosa}, meropenem resistance reached AUROC 0.834, but these agent-specific models should not be interpreted as direct validation of the composite IDSA difficult-to-treat-resistance phenotype. MRSA prediction reached AUROC 0.701, and VRE models reached 0.700 in \emph{E. faecium} and 0.812 in \emph{E. faecalis}. These conditional estimates support organism-specific prediction after organism identification, but do not establish the performance of the full pre-culture hierarchy.

\begin{table}[htbp]
\captionsetup{position=top}
\caption{\textbf{Representative organism-specific resistance-phenotype model performance.} Models were pooled across sites, specimen types and care settings. Stanford, ECUH and UTSW used chronological partitions; MGB used a patient-hash split because timestamps were shifted. Performance is conditional on the observed organism and does not measure the end-to-end organism-prediction hierarchy. IDSA section labels locate the corresponding guideline sections; the P. aeruginosa row is agent-specific rather than a composite difficult-to-treat-resistance phenotype, and MRSA and VRE are Gram-positive complements. Section abbreviations are ESBL-producing Enterobacterales (ESBL-E) and AmpC \ensuremath{\beta}-lactamase-producing Enterobacterales (AmpC-E).}\label{tab:2}
\centering\footnotesize
\begin{tabular}{@{}lllrr@{}}
\toprule
\textbf{IDSA Section} & \textbf{Organism} & \textbf{Phenotype} & \textbf{AUROC} & \textbf{Test n} \\
\midrule
section 1 ESBL-E & \emph{E. coli} & Ceftriaxone R & 0.759 & 46,752 \\
section 1 ESBL-E & \emph{K. pneumoniae} & Ceftriaxone R & 0.793 & 11,250 \\
section 2 AmpC-E & \emph{E. cloacae} & Cefepime R & 0.689 & 1,569 \\
section 3 CRE & \emph{Klebsiella} spp. & Ertapenem R & 0.676 & 13,965 \\
section 4 & \emph{P. aeruginosa} & Meropenem R & 0.834 & 5,693 \\
Gram+ & \emph{S. aureus} & MRSA & 0.701 & 5,848 \\
Gram+ & \emph{E. faecium} & VRE & 0.700 & 756 \\
\bottomrule
\end{tabular}
\end{table}

\subsection{Drivers of model predictions}\label{sec:drivers-of-model-predictions}

Across the 23 released empiric models, a prior-resistance feature ranked first in 13 and among the top two in 18 (Extended Data Fig.~\ref{edfig:6}). Leading prior-resistance variables included event counts, same-antibiotic resistance within one year and time since the most recent resistance event. Recent antibiotic exposure ranked among the ten leading encoded features in 19 models and an ADI feature in 15. Representative Gram-stage and resistance-phenotype models showed similar contributions from prior infection and antibiotic exposure. Shapley additive explanations (SHAP) describe associations with model output. They do not establish causal mechanisms or exclude proxying of site and care processes.

\section{Discussion}\label{sec:discussion}

ARMD-Predict combines two complementary approaches to empiric antibiotic selection using routinely collected EHR data. Across four academic health systems, pooled models showed broadly consistent discrimination across contributing sites. The principal comparison with a transported Stanford-only model, however, used different evaluation rows and cannot support a patient-matched estimate of the gain from pooling. Adding site identity and GroupDRO did not improve on simple pooling, and the released models contain no site feature. Susceptibility models had low reported calibration error, although selection of the calibration transform reused the test set. The organism-specific phenotype models were evaluated conditional on the observed organism rather than as an end-to-end hierarchy. Exploratory operating points identified isolate observations for narrowing or broader coverage, and prior resistance and recent antibiotic exposure were leading predictive features.

Prior patient-specific resistance studies focused on urinary tract infection, outpatient prescribing and personalized antibiograms\cite{yelin2019personal,kanjilal2020decision,corbin2020personalized,corbin2022personalized}. The present study adds pooled training across four sites, an organism-and-resistance hierarchy and explicit probability calibration. Matrix-assisted laser desorption/ionization time-of-flight methods can support cross-hospital resistance prediction, but require laboratory measurements after specimen processing and address a later stage of care\cite{weis2022maldi}. In contrast, the intended short-form ARMD-Predict models use information available around culture ordering. The two components also answer different questions. The empiric model estimates activity of a candidate antibiotic against a recovered isolate, whereas the hierarchical model decomposes the pre-culture problem into Gram class, organism identity and organism-specific phenotype. Across ten completed organism-classifier variants, top-3 accuracy ranged from 51.3\% to 83.3\%; four variants retained duplicate organism labels that divided probability across classes. Good phenotype discrimination given the true organism does not validate organism identification and phenotype prediction as one clinical pipeline.

The transportability results support multi-site pooling as a practical basis for model development within the represented health systems, but not as a guarantee of external performance. Performance of the Stanford-only baseline changed unevenly across sites, and the available pooled comparison did not evaluate the same patients. Chronological retraining reduced AUROC by approximately 0.04, suggesting that same-period patient-grouped estimates are optimistic for future use. All participating institutions were U.S. academic medical centers, and performance in community hospitals and international settings is unknown. Independent temporal and geographic validation, using fixed models and patient-matched comparisons, is therefore needed before attributing improved transportability to pooling.

Probability calibration requires a separate validation strategy from discrimination. The pooled models had low reported ECE, whereas raw Stanford-only outputs were overconfident and improved after isotonic transformation. However, choosing between the raw output and candidate transforms using test-set ECE and then reporting that same ECE makes the final estimate optimistic. Local-prevalence re-anchoring reduced ECE most when deployment prevalence differed from training, but the analysis used the held-out site's realized prevalence. A prospective site should prespecify the historical antibiogram window, lock the transform and prevalence estimate before evaluation, and monitor calibration level and slope as resistance patterns change.

The two-threshold analysis links model output to projected prescribing decisions. For ceftriaxone in inpatient urine cultures, an exploratory threshold with 95\% positive predictive value identified 55\% of susceptible isolate observations for narrowing and flagged 60\% of resistant observations for broader coverage. Across the 17 models with a feasible operating point, 26--70\% of resistant observations were flagged, whereas the proportion eligible for narrowing ranged from 0.3\% to 100\%. Three cells had baseline susceptibility above 95\%, so the predictive-value target could be met without model discrimination, and six models never reached the target. The potential stewardship yield therefore depends on baseline susceptibility and ranking performance, and on clinician willingness to act on a model-generated susceptibility estimate. Because thresholds and yields were derived and evaluated on the same test rows, and polymicrobial cultures may contribute discordant organism-level labels, these estimates do not show that prescribing or patient outcomes would improve. They define candidate rules for validation with independent data and prospective workflow studies.

The feature and subgroup analyses add clinical context without establishing mechanism or equity. In representative models, prior resistance, recent antibiotic exposure and an ADI feature were among the leading encoded features. These associations are consistent with known resistance risk, but may also reflect site and care processes. Median within-model AUROC ranges were 0.02 by sex, 0.07 by age band and 0.05 by ADI quintile, with the widest differences in small subgroups. Similar discrimination across measured groups does not establish equitable recommendations or outcomes\cite{pfohl2024toolbox}. Race and ethnicity were unavailable, sex labels were inferred at three sites, and ADI is a neighborhood-level rather than individual measure of disadvantage\cite{kind2018making}.

The publicly accessible applications permit external technical evaluation, but remain research prototypes. The empiric component returns an antibiotic-specific probability and an experimental local-prevalence adjustment; the organism component returns ranked organism labels and organism-specific phenotype estimates. Neither application has been prospectively validated, integrated into an EHR or shown to change prescribing. Clinical evaluation should begin with calibration transforms and operating thresholds selected without test-set reuse, end-to-end evaluation of the organism hierarchy and usability testing. An initial silent prospective phase should lock thresholds in advance, quantify alert and override burden and measure order-level microbiological coverage before recommendations are shown. Later evaluation should assess time to active therapy, broad-spectrum exposure, adverse events, patient outcomes and subgroup effects. Model artifacts and analysis code can support independent technical assessment before clinical integration.

Several limitations concern the outcome definition. Intermediate results were grouped with resistant, and culture-negative or contaminated specimens were excluded. Polymicrobial cultures contributed one label per recovered organism rather than one order-level coverage outcome; 317,798 order--antibiotic pairs involved more than one organism, and 78,394 had discordant susceptibility. The analysis may therefore overstate coverage when an antibiotic covers one organism but misses another. Laboratory-inferred susceptibility labels accounted for 7.1--21.4\% of observations across sites. Two inferred-susceptibility extracts reached the spreadsheet row limit, leaving an unknown number of labels missing and potentially biasing analyses that depend on inferred results. The models also reflect historical testing and prescribing decisions and may reproduce associated confounding\cite{wong2021external,wiens2019no,nestor2019feature}.

Data provenance and evaluation design impose further constraints. Ordering mode was unavailable for 1.55\% of observations, and local encounter mappings prevented separate evaluation of emergency and observation settings. Laboratory and vital-sign summaries could include same-day information that did not precede the culture order, so only the short-form feature set can be interpreted as reliably available at that decision point. Antibiotic names were standardized using a manually curated alias map rather than a formal drug ontology; residual errors could split or merge an agent's data across sites. Organism-model calendar splits may allow patients with cultures spanning a cutoff to contribute to more than one partition. The local-prevalence adjustment assumes prior-probability shifts rather than changes in predictor--outcome relationships, and its retrospective evaluation used each held-out site's realized prevalence, stronger information than a prospective hospital would ordinarily possess. Prospective assessment should use a prespecified historical prevalence estimate, monitor both calibration level and slope, and test whether performance remains stable as resistance epidemiology and clinical practice change.

\section{Conclusion}\label{sec:conclusion}

ARMD-Predict provides EHR-based estimates of isolate susceptibility, organism identity and organism-specific resistance conditional on the observed organism. Multi-site training may improve performance consistency within the represented health systems, and the retrospective operating-point analyses generate testable stewardship hypotheses. Independent calibration and threshold validation, end-to-end evaluation of the organism hierarchy, complete label-source auditing and prospective studies remain necessary. While further testing is needed, these results suggest that personalized antibiograms drawing on a broad array of routinely collected variables may help guide empiric treatment decisions.

\section*{Data and code availability}

Both research applications are available at \url{https://armd-predict.stanford.edu}.

Analysis code, model artifacts and the calibration pipeline are available at \url{https://github.com/HealthRex/CDSS/tree/master/scripts/antibiotic-susceptibility/armd-predict}. Access to site-level clinical data is governed by institutional approvals and data-use agreements.

\section*{Acknowledgments}
We thank the data-engineering and informatics teams at Stanford Health Care, ECUH, UTSW and MGB for assembling and harmonizing the de-identified EHR extracts. We thank the antimicrobial stewardship and infectious diseases teams at each site for guidance on prescribing logic and resistance-phenotype definitions.

\bibliographystyle{unsrtnat}
\bibliography{references}

\clearpage
\section*{Extended Data}
\setcounter{figure}{0}
\setcounter{table}{0}
\captionsetup[figure]{name={Extended Data Fig.}}
\captionsetup[table]{name={Extended Data Table}}
\renewcommand{\theHfigure}{ed.\arabic{figure}}
\renewcommand{\theHtable}{ed.\arabic{table}}
\begin{figure}[htbp]
\centering
\includegraphics[width=\textwidth]{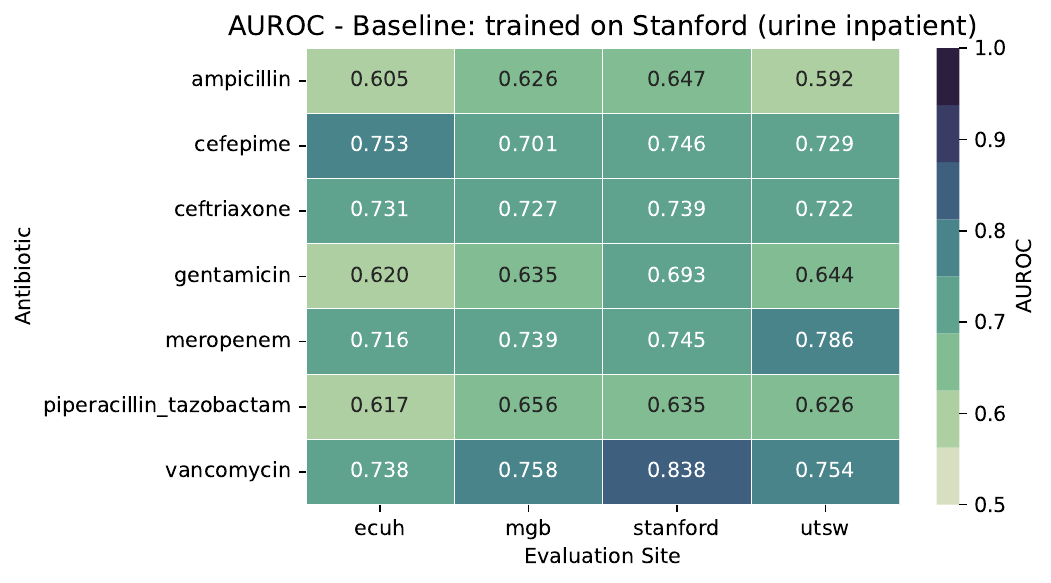}
\caption{\textbf{Transportability of Stanford-only inpatient urine models.} Models were evaluated on Stanford's held-out test subset and on the complete cohorts of three hospitals excluded from training. AUROC varied across sites; the largest decrease in this panel was for vancomycin at ECUH (0.838 to 0.738). The unequal evaluation scopes mean that comparisons with the pooled model in Fig.~\ref{fig:3} are descriptive rather than patient-matched.}
\label{edfig:1}
\end{figure}

\begin{figure}[htbp]
\centering
\includegraphics[width=0.85\textwidth]{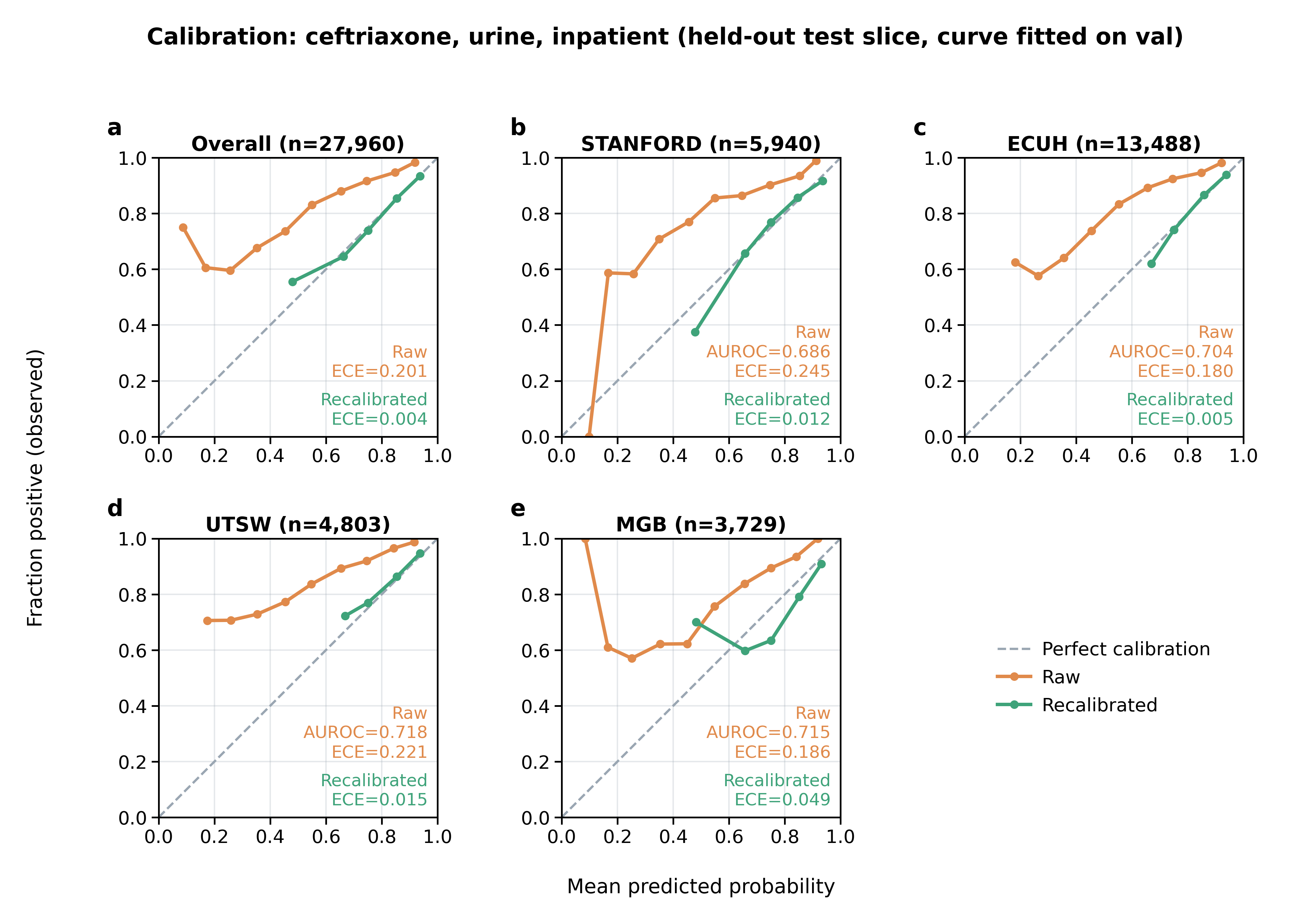}
\caption{\textbf{Probability calibration before and after isotonic recalibration.} Reliability diagrams show a representative ceftriaxone inpatient urine model overall and by site; the diagonal denotes perfect calibration, orange curves are raw XGBoost estimates and green curves follow isotonic transformation. Candidate transforms were fitted on validation data, but transform acceptance in the current analysis used test-set ECE. The curves and reported ECE values are therefore descriptive post-selection estimates and should be re-estimated after validation-only model selection.}
\label{edfig:2}
\end{figure}

\begin{figure}[htbp]
\centering
\includegraphics[width=\textwidth]{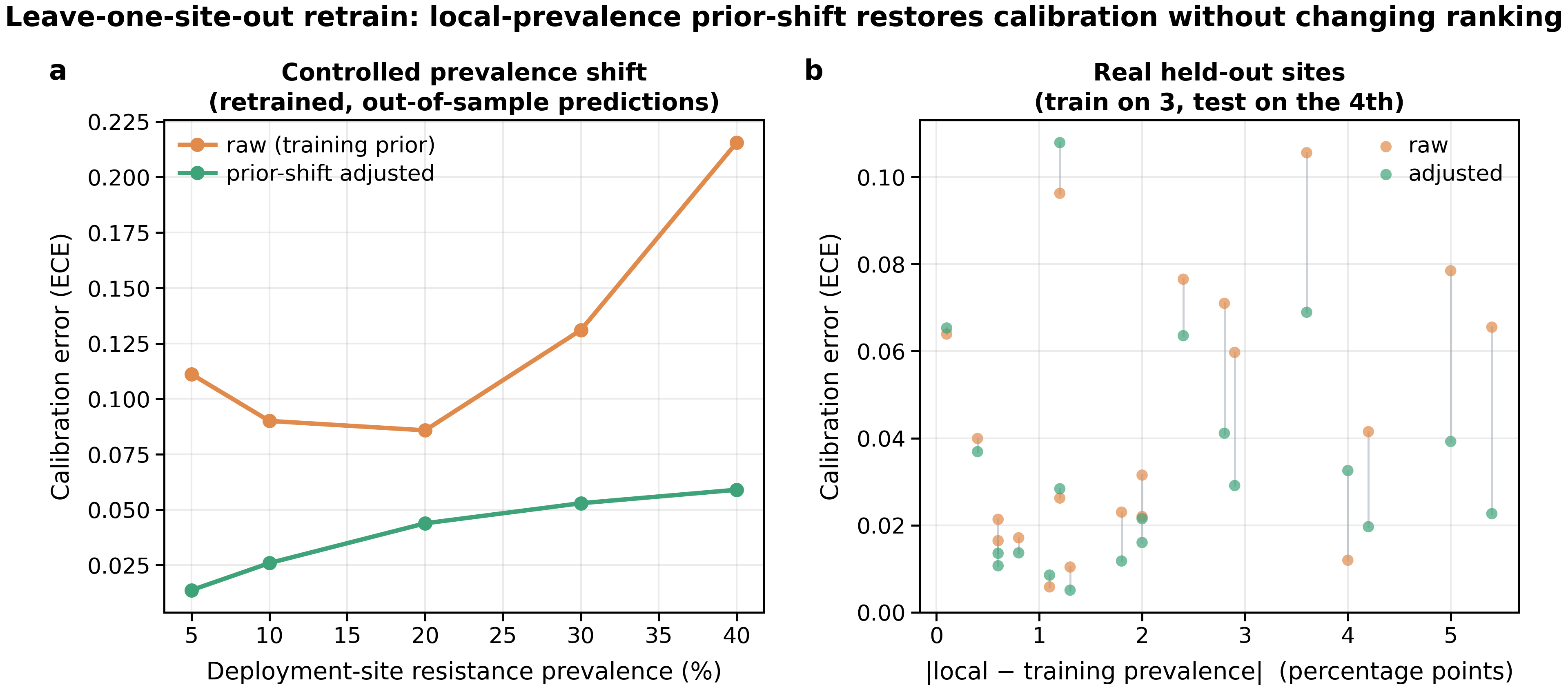}
\caption{\textbf{Leave-one-site-out evaluation of local-prevalence re-anchoring.} Models were trained on three sites, recalibrated on a held-out training slice and evaluated at the fourth. A prior-probability adjustment re-anchored predictions to the held-out site's realized resistance prevalence. In a controlled prevalence sweep, the adjustment reduced ECE as deployment prevalence diverged from training; on the observed held-out sites, gains were largest where prevalence differed most. The AUROC was unchanged because the transformation was monotonic. Because the retrospective analysis used realized held-out-site prevalence, it represents an oracle-prevalence implementation rather than a prospective historical-antibiogram evaluation.}
\label{edfig:3}
\end{figure}

\begin{figure}[htbp]
\centering
\includegraphics[width=\textwidth]{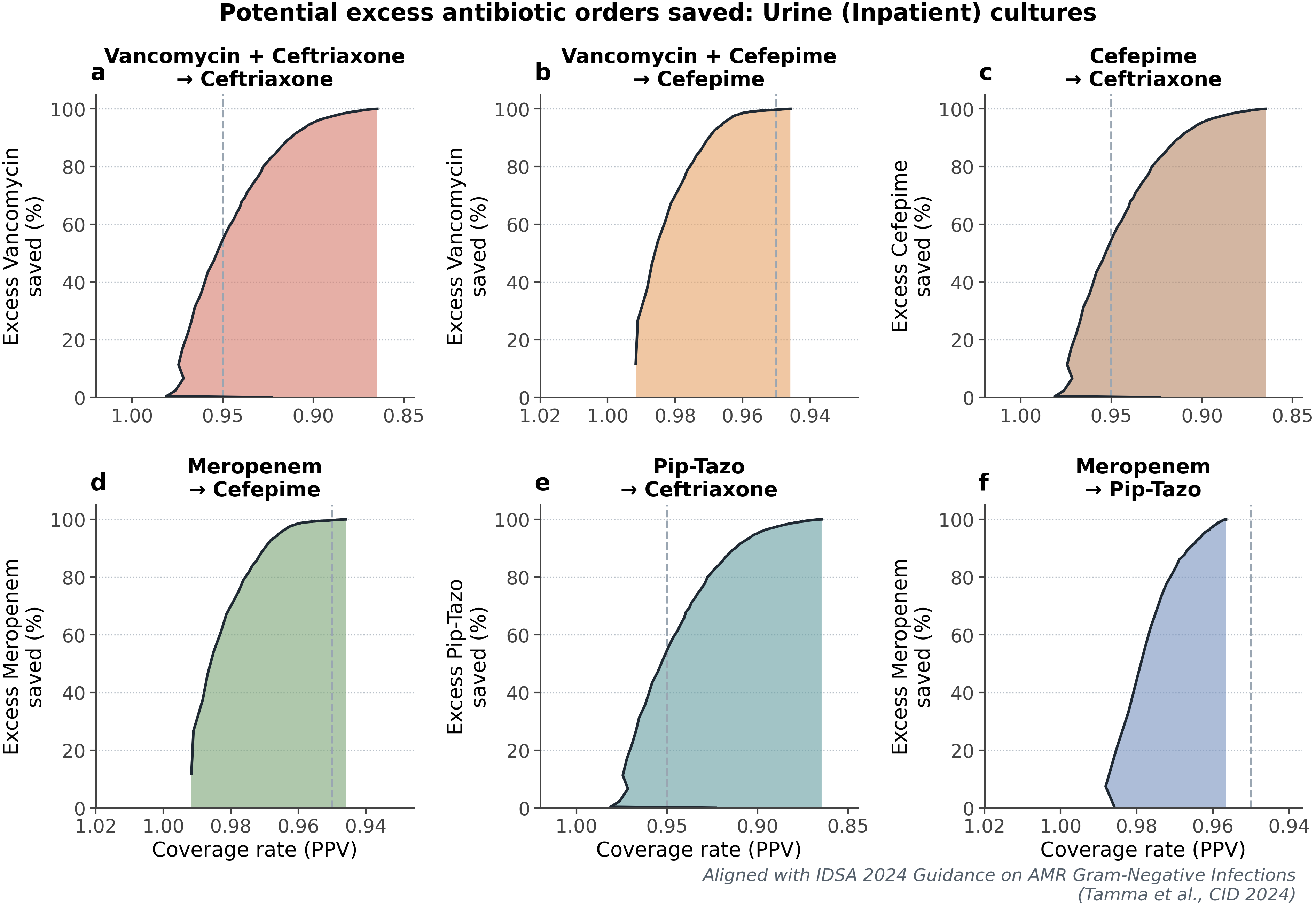}
\caption{\textbf{Projected narrowing envelopes for inpatient urine cultures.} Each panel represents a broad-to-narrow scenario, and the y axis names the broader agent whose use would be avoided under assumed adherence. Curves relate the positive predictive value of a narrowing recommendation to the proportion of susceptible isolate observations identified; panels sharing a narrow agent therefore share a curve. The dashed line marks 95\% positive predictive value. Curves and operating points were derived from the same test observations and do not estimate actual orders avoided. The corresponding blood-culture analysis is shown in Extended Data Fig.~\ref{edfig:5}.}
\label{edfig:4}
\end{figure}

\begin{figure}[htbp]
\centering
\includegraphics[width=0.8\textwidth]{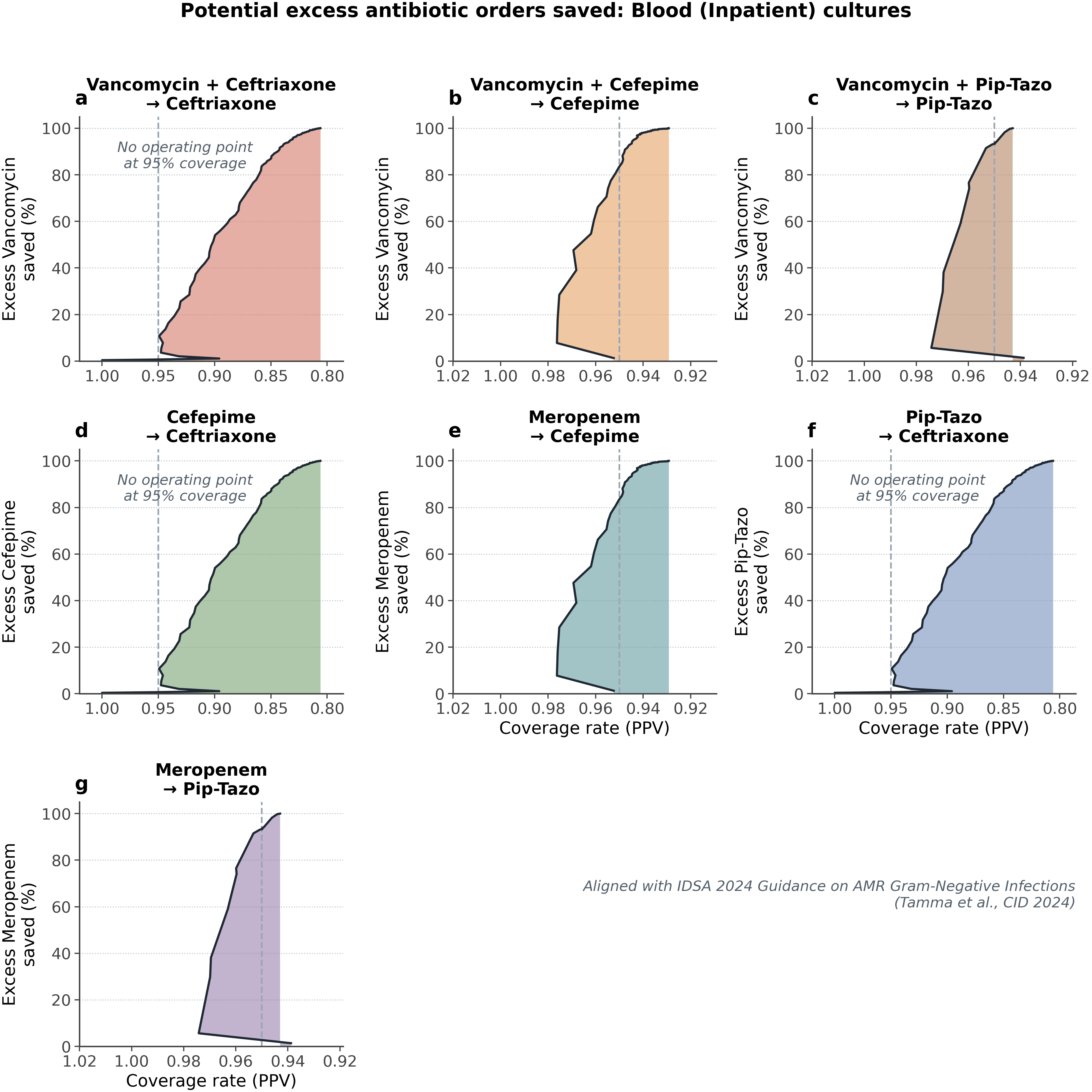}
\caption{\textbf{Projected narrowing envelopes for inpatient blood cultures.} The seven broad-to-narrow scenarios use the analysis described for Extended Data Fig.~\ref{edfig:4}. Panels sharing a narrow agent share a curve, read against different broader agents. Urine and blood analyses used different observations and independently scaled axes, so their comparison is descriptive. The dashed line marks 95\% positive predictive value.}
\label{edfig:5}
\end{figure}

\begin{figure}[htbp]
\centering
\includegraphics[width=\textwidth]{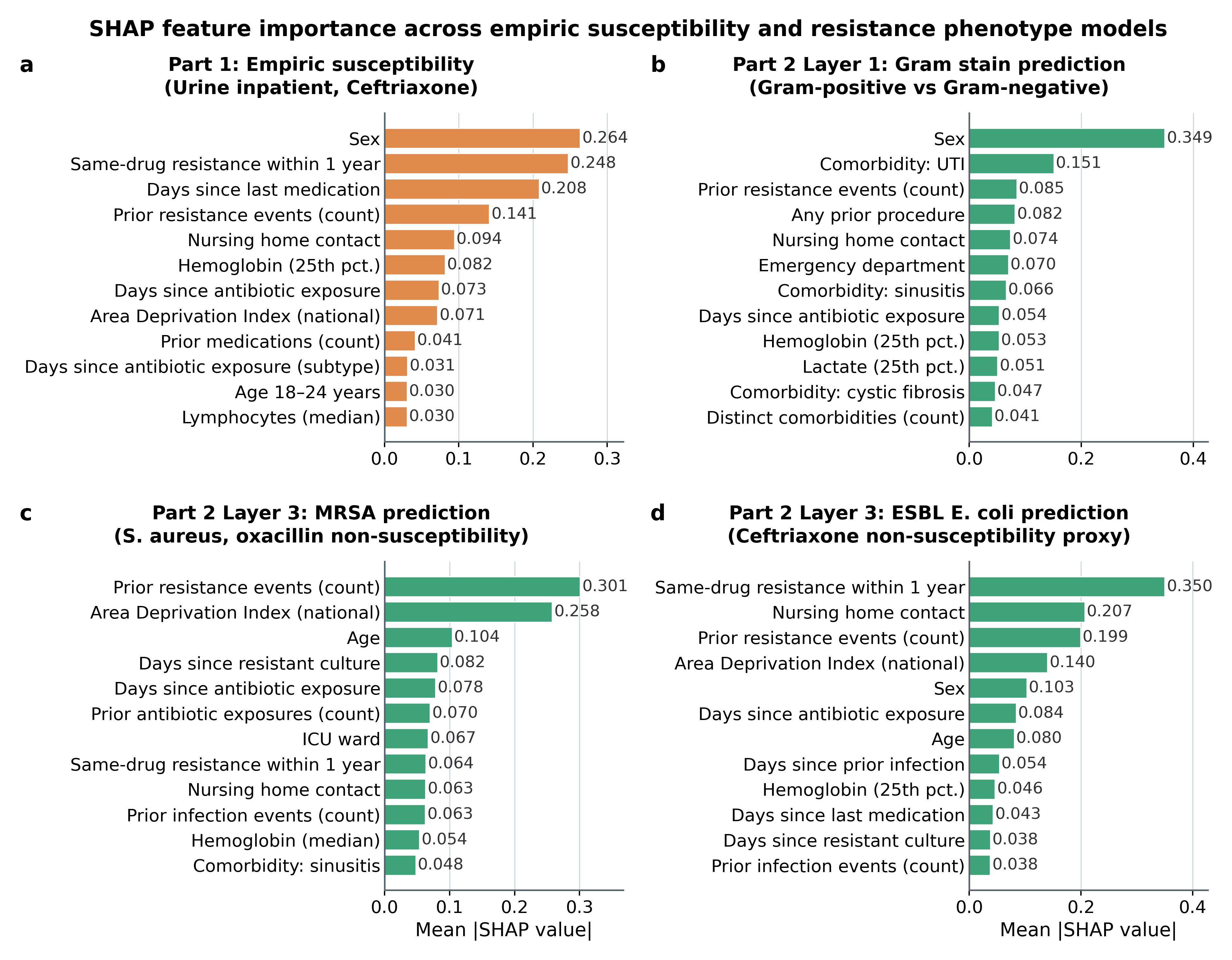}
\caption{\textbf{Feature attribution across representative models.} Mean absolute SHAP values were computed on held-out test observations for an empiric ceftriaxone inpatient urine model, Gram-stain classification, MRSA prediction using oxacillin non-susceptibility and an ESBL proxy using E. coli ceftriaxone non-susceptibility. Prior-resistance features ranked highly in the empiric and phenotype models; ADI and prior antibiotic exposure also appeared among leading features. SHAP attribution describes associations with model output and does not establish a causal mechanism or exclude proxying of site or care processes.}
\label{edfig:6}
\end{figure}

\begin{figure}[htbp]
\centering
\includegraphics[width=\textwidth]{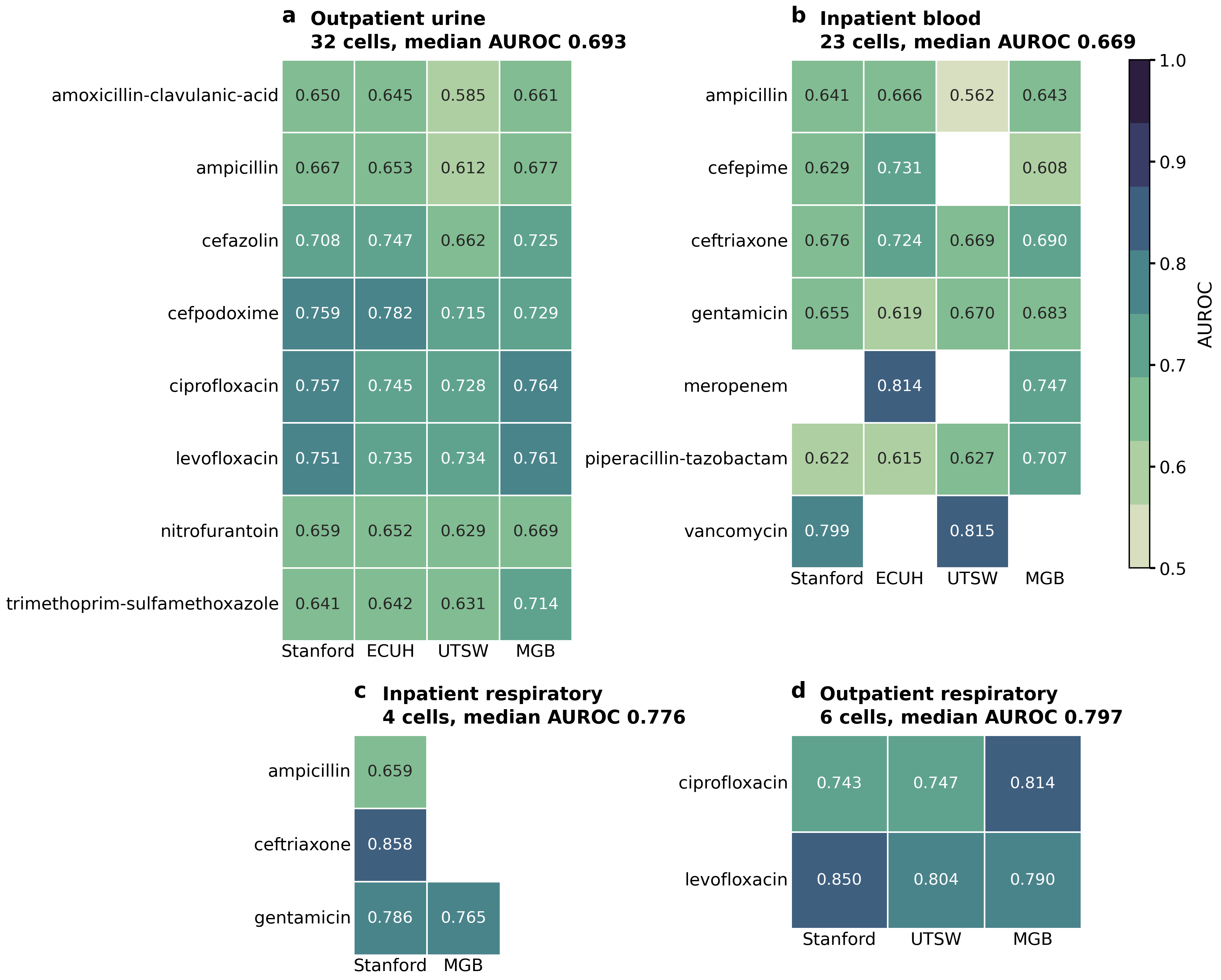}
\caption{\textbf{Discrimination of the site-aware pooled model in the remaining specimen and setting strata.} Heatmaps report held-out AUROC for each antibiotic and evaluation site, using the analysis described for Fig. 3, which shows inpatient urine. a, Outpatient urine (32 antibiotic--site cells; median AUROC 0.693). b, Inpatient blood (23 cells; median 0.669). c, Inpatient respiratory (4 cells; median 0.776). d, Outpatient respiratory (6 cells; median 0.797). Colour scale and palette are shared with Fig. 3 and span AUROC 0.5 to 1.0. Blank cells are antibiotic--site combinations in which that site's own held-out test rows did not meet the inclusion thresholds of at least 500 results with at least 25 susceptible and at least 25 resistant. The same thresholds were applied to the pooled data and independently to each of the training, validation and test partitions before an antibiotic was modelled at all, which is why the respiratory panels are sparse. Every hospital contributed training data and was evaluated on its own share of the held-out split.}
\label{edfig:7}
\end{figure}

\clearpage

\begingroup\scriptsize\setlength{\tabcolsep}{4pt}
\begin{xltabular}{\textwidth}{@{}Xrrrrr@{}}
\caption{\textbf{Cohort characteristics by study site.} Patient counts, culture orders, age, sex and ADI are summarized at each patient's earliest culture order; care setting, specimen and susceptibility are observation-level measures. Percentages therefore use different denominators. Patients are not deduplicated across health systems. Age was released in bands, precluding a median. Sex was released as an unlabeled binary variable at three sites; the displayed assignment was inferred from culture distributions and corroborated at MGB. ADI is a national percentile at Stanford, ECUH and UTSW, but a continuous ZIP-code mean at MGB and is not directly comparable across all columns. MGB dates were shifted during de-identification. Race and ethnicity were unavailable. Sex is reported as percentages so that categories with fewer than 20 patients cannot be recovered from the totals.}\label{edtab:1}\\
\toprule
\textbf{Characteristic} & \textbf{Stanford} & \textbf{ECUH} & \textbf{UTSW} & \textbf{MGB} & \textbf{Overall} \\
\midrule
\endfirsthead
\multicolumn{6}{@{}l}{\textit{(continued)}}\\
\toprule
\textbf{Characteristic} & \textbf{Stanford} & \textbf{ECUH} & \textbf{UTSW} & \textbf{MGB} & \textbf{Overall} \\
\midrule
\endhead
\bottomrule
\endlastfoot
Patients, n & 64,677 & 62,354 & 45,194 & 69,002 & 241,227 \\
Culture orders, n & 112,508 & 105,725 & 84,434 & 113,285 & 415,952 \\
Culture--antibiotic observations, n & 1,900,770 & 1,953,562 & 1,340,348 & 2,205,103 & 7,399,783 \\
Culture dates & 2007--2024 & 2014--2025 & 2006--2025 & - & 2006--2025 \\
\addlinespace[3pt]
\multicolumn{6}{@{}l}{Age band, n (\%)} \\
18--24 & 5,780 (8.9\%) & 6,219 (10.0\%) & 2,824 (6.2\%) & 2,881 (4.2\%) & 17,704 (7.3\%) \\
25--34 & 7,907 (12.2\%) & 4,531 (7.3\%) & 4,803 (10.6\%) & 6,724 (9.7\%) & 23,965 (9.9\%) \\
35--44 & 6,745 (10.4\%) & 6,465 (10.4\%) & 4,449 (9.8\%) & 5,979 (8.7\%) & 23,638 (9.8\%) \\
45--54 & 7,734 (12.0\%) & 6,315 (10.1\%) & 5,664 (12.5\%) & 6,620 (9.6\%) & 26,333 (10.9\%) \\
55--64 & 9,903 (15.3\%) & 8,796 (14.1\%) & 8,168 (18.1\%) & 10,712 (15.5\%) & 37,579 (15.6\%) \\
65--74 & 11,645 (18.0\%) & 11,644 (18.7\%) & 10,095 (22.3\%) & 14,631 (21.2\%) & 48,015 (19.9\%) \\
75--84 & 9,408 (14.5\%) & 10,474 (16.8\%) & 6,668 (14.8\%) & 12,928 (18.7\%) & 39,478 (16.4\%) \\
85--89 & 3,228 (5.0\%) & 4,336 (7.0\%) & 1,646 (3.6\%) & 4,483 (6.5\%) & 13,693 (5.7\%) \\
90 or above & 2,327 (3.6\%) & 3,574 (5.7\%) & 877 (1.9\%) & 4,044 (5.9\%) & 10,822 (4.5\%) \\
\addlinespace[3pt]
\multicolumn{6}{@{}l}{Sex, \%} \\
Female & 73.7 & 69.8 & 70.7 & 71.8 & 71.6 \\
Male & 26.2 & 30.2 & 29.3 & 28.1 & 28.4 \\
Other or not recorded & $<$0.1 & 0 & $<$0.1 & $<$0.1 & $<$0.1 \\
\addlinespace[3pt]
\multicolumn{6}{@{}l}{ADI} \\
Median (Q1--Q3) & 3 (1--7) & 80 (70--89) & 49 (31--66) & 18 (10--26) & - \\
Not recorded, n (\%) & 14,630 (22.6\%) & 3,170 (5.1\%) & 2,028 (4.5\%) & 30,770 (44.6\%) & 50,598 (21.0\%) \\
\addlinespace[3pt]
\multicolumn{6}{@{}l}{Care setting, n (\%)} \\
Inpatient & 1,015,867 (53.4\%) & 1,630,795 (83.5\%) & 691,090 (51.6\%) & 708,613 (32.1\%) & 4,046,365 (54.7\%) \\
Outpatient & 884,903 (46.6\%) & 322,767 (16.5\%) & 649,258 (48.4\%) & 1,496,490 (67.9\%) & 3,353,418 (45.3\%) \\
\addlinespace[3pt]
\multicolumn{6}{@{}l}{Culture type, n (\%)} \\
Urine & 1,549,804 (81.5\%) & 1,669,723 (85.5\%) & 1,107,775 (82.6\%) & 1,877,095 (85.1\%) & 6,204,397 (83.8\%) \\
Blood & 158,917 (8.4\%) & 223,849 (11.5\%) & 74,673 (5.6\%) & 185,038 (8.4\%) & 642,477 (8.7\%) \\
Respiratory & 192,049 (10.1\%) & 59,990 (3.1\%) & 157,900 (11.8\%) & 142,970 (6.5\%) & 552,909 (7.5\%) \\
Susceptible results, n (\%) & 1,492,737 (78.5\%) & 1,606,122 (82.2\%) & 1,036,658 (77.3\%) & 1,773,249 (80.4\%) & 5,908,766 (79.9\%) \\
\end{xltabular}
\endgroup

\begingroup\footnotesize
\begin{longtable}{@{}llrrrl@{}}
\caption{\textbf{Resistance prevalence across the four study sites compared with national U.S. CDC and NHSN estimates from 2022.}}\label{edtab:2}\\
\toprule
\textbf{Organism} & \textbf{Antibiotic} & \textbf{Isolates (n)} & \textbf{Our sites (\%)} & \textbf{National (\%)} & \textbf{Source} \\
\midrule
\endfirsthead
\multicolumn{6}{@{}l}{\textit{(continued)}}\\
\toprule
\textbf{Organism} & \textbf{Antibiotic} & \textbf{Isolates (n)} & \textbf{Our sites (\%)} & \textbf{National (\%)} & \textbf{Source} \\
\midrule
\endhead
\bottomrule
\endlastfoot
\emph{S. aureus} & Oxacillin (MRSA) & 28,251 & 34.0 & $\sim$43 & CDC/NHSN 2022 \\
\emph{E. faecium} & Vancomycin (VRE) & 2,692 & 54.3 & $\sim$30 & CDC/NHSN 2022 \\
\emph{E. coli} & Fluoroquinolone & 198,216 & 24.5 & $\sim$28 & CDC/NHSN 2022 \\
\end{longtable}
\endgroup

\begingroup\footnotesize
\begin{xltabular}{\textwidth}{@{}Xrrrr@{}}
\caption{\textbf{Subgroup discrimination audit.} For each released antibiotic--specimen model, the table reports the maximum minus minimum test-set AUROC across sex, age band and ADI quintile. Median within-model ranges were 0.02, 0.07 and 0.05, respectively. CIs used 1,000 class-stratified bootstrap replicates. These analyses assess discrimination only and do not establish calibration, operating-threshold parity or equitable clinical effects. Sex labels were inferred at three sites, and the bootstrap sampled rows rather than patients.}\label{edtab:3}\\
\toprule
\textbf{Antibiotic (specimen/setting)} & \textbf{Overall AUROC} & \textbf{Sex} & \textbf{Age band} & \textbf{ADI} \\
\midrule
\endfirsthead
\multicolumn{5}{@{}l}{\textit{(continued)}}\\
\toprule
\textbf{Antibiotic (specimen/setting)} & \textbf{Overall AUROC} & \textbf{Sex} & \textbf{Age band} & \textbf{ADI} \\
\midrule
\endhead
\bottomrule
\endlastfoot
amoxicillin\_clavulanic\_acid (urine/outpatient) & 0.638 & 0.001 & 0.025 & 0.038 \\
ampicillin (blood/inpatient) & 0.636 & 0.010 & 0.037 & 0.055 \\
ampicillin (urine/inpatient) & 0.629 & 0.011 & 0.025 & 0.030 \\
cefazolin (urine/outpatient) & 0.708 & 0.021 & 0.058 & 0.027 \\
cefepime (blood/inpatient) & 0.704 & 0.004 & 0.071 & 0.129 \\
cefepime (urine/inpatient) & 0.807 & 0.038 & 0.071 & 0.040 \\
cefpodoxime (urine/outpatient) & 0.733 & 0.050 & 0.016 & 0.060 \\
ceftriaxone (blood/inpatient) & 0.703 & 0.028 & 0.040 & 0.078 \\
ceftriaxone (resp/inpatient) & 0.825 & 0.019 & 0.175 & 0.034 \\
ceftriaxone (urine/inpatient) & 0.773 & 0.043 & 0.041 & 0.028 \\
ciprofloxacin (urine/outpatient) & 0.743 & 0.053 & 0.027 & 0.047 \\
gentamicin (blood/inpatient) & 0.656 & 0.010 & 0.100 & 0.059 \\
gentamicin (urine/inpatient) & 0.679 & 0.041 & 0.021 & 0.036 \\
levofloxacin (resp/outpatient) & 0.822 & 0.014 & 0.090 & 0.047 \\
levofloxacin (urine/outpatient) & 0.747 & 0.031 & 0.044 & 0.042 \\
meropenem (blood/inpatient) & 0.835 & 0.042 & 0.155 & 0.103 \\
meropenem (urine/inpatient) & 0.831 & 0.063 & 0.117 & 0.087 \\
nitrofurantoin (urine/outpatient) & 0.662 & 0.015 & 0.012 & 0.042 \\
piperacillin\_tazobactam (blood/inpatient) & 0.658 & 0.019 & 0.164 & 0.051 \\
piperacillin\_tazobactam (urine/inpatient) & 0.699 & 0.010 & 0.081 & 0.099 \\
trimethoprim\_sulfamethoxazole (urine/outpatient) & 0.673 & 0.040 & 0.072 & 0.062 \\
vancomycin (blood/inpatient) & 0.825 & 0.069 & 0.279 & 0.096 \\
vancomycin (urine/inpatient) & 0.825 & 0.011 & 0.075 & 0.044 \\
\end{xltabular}
\endgroup

\begingroup\footnotesize
\begin{xltabular}{\textwidth}{@{}llXrr@{}}
\caption{\textbf{Complete organism-specific resistance-phenotype model set.} The table reports the held-out-test AUROC for all 30 phenotype classifiers. IDSA section labels locate the corresponding guideline sections. P. aeruginosa rows are agent-specific models and do not constitute the composite difficult-to-treat-resistance phenotype. Performance is conditional on the observed organism rather than the predicted organism. The seven representative phenotypes in Table~\ref{tab:2} are a subset.}\label{edtab:4}\\
\toprule
\textbf{Organism} & \textbf{Antibiotic} & \textbf{Phenotype (IDSA)} & \textbf{AUROC} & \textbf{n} \\
\midrule
\endfirsthead
\multicolumn{5}{@{}l}{\textit{(continued)}}\\
\toprule
\textbf{Organism} & \textbf{Antibiotic} & \textbf{Phenotype (IDSA)} & \textbf{AUROC} & \textbf{n} \\
\midrule
\endhead
\bottomrule
\endlastfoot
\emph{C. freundii} & ceftriaxone & AmpC proxy (section 2) & 0.689 & 748 \\
\emph{E. cloacae} & cefepime & AmpC proxy (section 2) & 0.689 & 1,569 \\
\emph{E. cloacae} & ceftriaxone & AmpC proxy (section 2) & 0.698 & 1,184 \\
\emph{E. coli} & TMP-SMX & TMP-SMX R & 0.653 & 48,290 \\
\emph{E. coli} & ceftriaxone & ESBL proxy (section 1) & 0.759 & 46,752 \\
\emph{E. coli} & ciprofloxacin & Fluoroquinolone R & 0.746 & 46,904 \\
\emph{E. coli} & ertapenem & CRE proxy (section 3) & 0.691 & 43,982 \\
\emph{E. coli} & nitrofurantoin & Nitrofurantoin & 0.662 & 45,496 \\
\emph{E. faecalis} & ampicillin & Ampicillin & 0.519 & 5,795 \\
\emph{E. faecalis} & vancomycin & VRE (vancomycin) & 0.812 & 5,805 \\
\emph{E. faecium} & ampicillin & Ampicillin & 0.752 & 754 \\
\emph{E. faecium} & linezolid & Linezolid & 0.634 & 657 \\
\emph{E. faecium} & vancomycin & VRE (vancomycin) & 0.700 & 756 \\
\emph{Enterobacter} spp. & meropenem & CRE proxy (section 3) & 0.734 & 1,710 \\
\emph{Enterococcus} spp. & vancomycin & VRE (vancomycin) & 0.812 & 6,957 \\
\emph{K. aerogenes} & ceftriaxone & AmpC proxy (section 2) & 0.695 & 1,048 \\
\emph{K. oxytoca} & ceftriaxone & ESBL proxy (section 1) & 0.687 & 1,349 \\
\emph{K. pneumoniae} & ceftriaxone & ESBL proxy (section 1) & 0.793 & 11,250 \\
\emph{K. pneumoniae} & ciprofloxacin & Fluoroquinolone R & 0.742 & 11,688 \\
\emph{Klebsiella} spp. & ertapenem & CRE proxy (section 3) & 0.676 & 13,965 \\
\emph{P. aeruginosa} & cefepime & \emph{P. aeruginosa}, agent-specific (section 4) & 0.790 & 5,884 \\
\emph{P. aeruginosa} & ceftazidime & \emph{P. aeruginosa}, agent-specific (section 4) & 0.751 & 5,955 \\
\emph{P. aeruginosa} & ciprofloxacin & \emph{P. aeruginosa}, agent-specific (section 4) & 0.759 & 5,859 \\
\emph{P. aeruginosa} & meropenem & \emph{P. aeruginosa}, agent-specific (section 4) & 0.834 & 5,693 \\
\emph{P. aeruginosa} & pip/tazo & \emph{P. aeruginosa}, agent-specific (section 4) & 0.725 & 5,857 \\
\emph{P. mirabilis} & ceftriaxone & ESBL proxy (section 1) & 0.719 & 5,112 \\
\emph{P. mirabilis} & ciprofloxacin & Fluoroquinolone R & 0.787 & 5,166 \\
\emph{S. aureus} & TMP-SMX & TMP-SMX susceptibility & 0.703 & 5,624 \\
\emph{S. aureus} & oxacillin & MRSA (oxacillin) & 0.701 & 5,848 \\
\emph{S. marcescens} & ceftriaxone & AmpC proxy (section 2) & 0.662 & 842 \\
\end{xltabular}
\endgroup

\begingroup\footnotesize
\begin{xltabular}{\textwidth}{@{}lrX@{}}
\caption{\textbf{Feature blocks retained in the pooled model.} Column counts are the union across specimen and setting strata after the all-site intersection and the automated leakage check. Categorical columns are one-hot expanded at training, so the fitted matrix is wider than the counts shown. The complete column list is provided as a supplementary file.}\label{edtab:5}\\
\toprule
\textbf{Feature block} & \textbf{Columns} & \textbf{What these variables capture} \\
\midrule
\endfirsthead
\multicolumn{3}{@{}l}{\textit{(continued)}}\\
\toprule
\textbf{Feature block} & \textbf{Columns} & \textbf{What these variables capture} \\
\midrule
\endhead
\bottomrule
\endlastfoot
Demographics & 2 & Age, grouped into nine bands from 18-24 years to 90 and over, and sex \\
Care setting & 6 & Where the culture was ordered: emergency department, intensive care, inpatient ward or outpatient clinic \\
Area Deprivation Index & 2 & Neighborhood socioeconomic deprivation at the patient's ZIP code, as a national percentile and a state decile \\
Prior resistance events & 7 & Whether the patient has had a resistant culture before, how long ago, how many, and whether any was to the antibiotic being predicted \\
Prior infecting organisms & 7 & Whether the patient has had a positive culture before, how long ago and how many. The organism identity is not carried forward \\
Prior antibiotic exposure, agent & 11 & Prior antibiotic courses: whether any, how recently, how many, and the specific agent most recently given \\
Prior antibiotic exposure, class & 9 & The same prior-exposure history summarized at antibiotic class level \\
Prior antibiotic exposure, subtype & 10 & The same prior-exposure history summarized at antibiotic subtype level \\
Prior invasive procedures & 7 & Prior invasive procedures, including dialysis, central venous catheters and mechanical ventilation: whether any, how recently and how many \\
Nursing-home contact & 2 & Whether the record shows contact with a nursing home \\
Comorbidities & 506 & One indicator per recorded condition, with the number of distinct conditions and whether any remained active at the time of culture. This is the only block in which several values can be present at once \\
Laboratory values & 56 & Lower and upper quartiles of eleven routine results: white cell count, neutrophils, lymphocytes, hemoglobin, platelets, creatinine, blood urea nitrogen, sodium, bicarbonate, lactate and procalcitonin \\
\textbf{Total} & \textbf{625} &  \\
\end{xltabular}
\endgroup

\end{document}